\documentclass[reprint,aps,prx,superscriptaddress]{revtex4-2}

\usepackage{hyperref}       % hyperlinks
\usepackage{url}            % simple URL typesetting
\usepackage{amsmath,amssymb}
\usepackage{mathrsfs}
\usepackage{booktabs}       % professional-quality tables
\usepackage{amsfonts}       % blackboard math symbols
\usepackage{multirow}
\usepackage{graphicx}
\usepackage{tikz}
\usepackage[caption=false]{subfig}
\usepackage{siunitx}
\usepackage[normalem]{ulem} 
\usepackage[export]{adjustbox}
\usepackage{xcolor}
\usepackage{array}[=2016-10-06]
\newcolumntype{K}[1]{>{\centering\arraybackslash}p{#1}}
\graphicspath{ {./figs/} } 
\newcommand*\circled[1]{\tikz[baseline=(char.base)]{
		\node[shape=circle,draw,inner sep=2pt] (char) {#1};}}

\begin{document}

\title{Microwave photon detection at parametric criticality}

\author{Kirill Petrovnin}
\affiliation{QTF  Centre  of  Excellence and InstituteQ, 
	Department of Applied Physics, Aalto University, FI-00076 Aalto, Finland\\}

\author{Jiaming Wang}
\affiliation{QTF  Centre  of  Excellence and InstituteQ, 
	Department of Applied Physics, Aalto University, FI-00076 Aalto, Finland\\}

\author{Michael Perelshtein}
\affiliation{QTF  Centre  of  Excellence and InstituteQ, 
Department of Applied Physics, Aalto University, FI-00076 Aalto, Finland\\}

\author{Pertti Hakonen}
\affiliation{QTF  Centre  of  Excellence and InstituteQ, 
Department of Applied Physics, Aalto University, FI-00076 Aalto, Finland\\}

\author{Gheorghe Sorin Paraoanu}
\affiliation{QTF  Centre  of  Excellence and InstituteQ, 
Department of Applied Physics, Aalto University, FI-00076 Aalto, Finland\\}

\begin{abstract}

The detection of microwave fields at single-photon power levels is a much sought-after technology, with
practical applications in nanoelectronics and quantum information science. Here we demonstrate a simple yet powerful criticality-enhanced method of  microwave photon detection by operating a magnetic-field tunable Kerr Josephson parametric amplifier at the border of a first-order phase transition and close to the critical point. We obtain a 73\% efficiency and a dark-count rate of 167 kHz, corresponding to a responsivity of $1.3 \times 10^{17}~\mathrm{W}^{-1}$ and noise-equivalent power of 3.28 zW/$\sqrt{\rm Hz}$. We verify the single-photon operation by extracting the 
Poissonian statistics of a coherent probe signal.

\end{abstract}
\maketitle

\section{Introduction}

Similarly to the way in which entanglement is regarded as a resource 
for enhancing metrological precision, systems displaying critical behavior -- with small variations of physical parameters leading to macroscopic changes \cite{huang2008statistical,tauberCriticalDynamicsField2014} -- can be employed for sensitive detection. For example, superconductors near the critical temperature can become normal conducting even if only tiny amounts of energy are absorbed, and the ensuing change in electrical resistance can be easily measured. This principle underlies the design of transition-edge sensors widely used in astronomy \cite{Irwin2005}. However, classical phase transitions respond only to changes in thermodynamic variables, which restricts the range of applications. In contrast, quantum phase transitions \cite{sondhi1997continuous} are sensitive to variations in the parameters of a Hamiltonian.  
In recent years a significant number of theoretical proposals have been advanced toward the investigation of quantum phase transitions for fundamental metrological research  \cite{Zanardi2008, Rams2018, Cai2021, Lu2022}, for the preparation of quantum resources such as highly-squeezed spin states and extractable work in quantum batteries \cite{Gietka2022, Paternostro2022}, and for applications in the
sensing of electromagnetic or mechanical degrees of freedom
\cite{Tsang2013, Guta2016, garbe2020critical, Roscilde2018, Plenio2022, dicandiaCriticalParametricQuantum2023, Gang2023}.

The parametrically driven quantum oscillator with Kerr nonlinearity and dissipation encapsulates in a quintessential way the physics of quantum phase transitions. 
Several paradigmatic models map to it: the Lipkin-Meshkov-Glick model via a Holstein-Primakoff transformation \cite{Vidal2004}, the quantum Rabi model via a Schrieffer-Wolff transformation \cite{Plenio2015}, the Dicke model in the limit of suppressed  spin-fluctuations \cite{Feshke2012}, and the $\mathcal{P}\mathcal{T}$ symmetry-breaking model in the mean-field description \cite{Baranger2021}. However, investigating these models experimentally is often challenging. For example, resolving the structure of the ground state by extracting  virtual photons in the Rabi model requires sophisticated protocols \cite{Pino2023}, while for the parametrically driven harmonic oscillator this is straightforward. Moreover, using finite-component criticality instead of a many-body system presents multiple advantages such as simplicity of the device and Hamiltonian description, avoidance of non-local operations, and direct applicability of quantum control methods.

Here we employ a Kerr Josephson parametric amplifier (K-JPA) realized as a quarter-wavelength superconducting stripe terminated by a SQUID to map in detail the phase diagram and to detect single-photon microwave radiation. The effective electrical length can be tuned fast by using an external magnetic field, allowing us to parametrically pump the device at a frequency close to double of the resonant frequency, inducing 3-wave mixing. Similar devices have been studied in other contexts: bifurcation amplification \cite{vijay2009invited}, dynamical Casimir effect \cite{wilson2011observation}, generation of entanglement and cluster states \cite{PhysRevApplied.18.024063,https://doi.org/10.1002/qute.202200031}, qubit-readout protocols \cite{krantzSingleshotReadoutSuperconducting2016}, and unconventional forms of quantum computing by using qubits encoded in continuous variables \cite{grimm2020stabilization,puri2017engineering,berdou2023one,he2023fast,lu2023resolving}. However, while the behavior of this system below the parametric threshold is well understood and verified experimentally, the physics near and above the threshold has been less explored from the perspective of applications in sensing.
In our experiments, we operate the device as a criticality-enhanced detector by exploiting the crossing of a first-order phase transition triggered  by  incident quanta.

Our experiments demonstrate conclusively a new concept for a microwave single-photon detector (SPD) -- a much sought-after device in mesoscopic physics and nanoelectronics \cite{Pekola2020}. In the optical domain, the energy carried by a photon is as low as $10^{-19}\,\unit{\joule}$ ($\sim1\,\unit{\eV}$), yet it is still possible to realize single-photon detectors (SPDs) based on photomultiplier tubes, avalanche photodiodes, quantum dots, and superconducting materials including transition edge sensors and nanowires \cite{buller2009single,hadfieldSinglephotonDetectorsOptical2009,migdallSinglephotonGenerationDetection2013}. However, for microwave signals with four-to-five orders of magnitude lower energy, single photon detection becomes extremely challenging \cite{sathyamoorthyDetectingItinerantSingle2016}. Indeed, the standard photodiode principle cannot be applied due to the low photon-to-electron conversion efficiency at these frequencies as well as due to the lack of semiconductors with very small gaps. Yet, progress has been made by employing semiconducting nanowire quantum dots coupled resonantly to a cavity, albeit the efficiency obtained so far is only 6\%~\cite{Maisi2021}. Also recent efforts in hybrid superconductor–graphene–superconductor bolometers has lead to energy resolution as low as 30 GHz \cite{Lee2020,Kokkoniemi2020}. Other modern microwave single-photon detection schemes and proposals are utilizing the switching  of a current-biased Josephson junction from zero to a finite voltage state through absorbing a photon \cite{chenMicrowavePhotonCounter2011,poudel2012quantum,walsh2017graphene,PhysRevApplied.7.014012,PhysRevApplied16014025}, the cross-Kerr interaction in a Josephson array embedded in a cavity  \cite{PhysRevApplied.15.034074}, and various mappings of the photon onto the state of a superconducting qubit \cite{inomata2016single,narla2016robust,besse2018single,konoQuantumNondemolitionDetection2018,lescanneIrreversibleQubitPhotonCoupling2020}. These detection schemes may also need to be heralded. While some of these schemes have been successfully demonstrated in the laboratory, the design and fabrication is often complicated, with considerable operational overhead and less reliable parameters (e.g. frequency and decoherence times can change over hours for superconducting qubits). In contrast, our detector is simple to fabricate, requires only an easy, bolometer-style operation protocol, and it is non-heralded. Since the design does not use qubit-type elements, the detector is not affected by quasiparticle poisoning \cite{catelaniQuasiparticleRelaxationSuperconducting2011, serniakHotNonequilibriumQuasiparticles2018, krantzQuantumEngineerGuide2019, deleonMaterialsChallengesOpportunities2021, siddiqiEngineeringHighcoherenceSuperconducting2021, catelaniUsingMaterialsQuasiparticle2022, mannilaSuperconductorFreeQuasiparticles2022} and can be operated in a stable way over large periods of time.

The results presented here further confirm previous theoretical models and experimental observations of the parametric phase transition in similar systems \cite{wustmannParametricResonanceTunable2013a, Krantz_2013, wilsonPhotonGenerationElectromagnetic2010}, and in particular they put forward the description utilizing a slow variable and its effective potential \cite{lin2015critical, marthaler2007quantum, peano2012sharp,dykmanFluctuatingNonlinearOscillators2012}.
This proof-of-concept detector for criticality-enhanced sensing demonstrates that experimental control over this effective potential is possible and may lead to a new class of devices and applications.
Indeed, in mesoscopic physics, the formation of effective potentials, typically by circuit engineering, is a key strategy for realizing devices such as superconducting qubits.

The realization of an efficient and robust single microwave photon detector will open up new opportunities in quantum information processing, metrology, and sensing. For example, the SPD is the key ingredient in one-way quantum computing \cite{walther2005experimental} and in quantum illumination with microwave photons \cite{sanz2017quantum}; it can be used for qubit readout \cite{zorin2011period,dicandiaCriticalParametricQuantum2023} and it would enable dark-matter axion detection with sensitivity below the standard quantum limit \cite{lamoreaux2013analysis,braine2020extended,dixit2021searching,kutluCharacterizationFluxdrivenJosephson2021}. 
	
The paper is organized as follows. We first introduce the device and obtain  the quantum phase diagram under parametric operation.  We identify the operational point near the border of the phase transition. Further, we describe the pulse sequence, we characterize the device as a detector, and we verify the Poissonian statistics of photons in a probe field.

\section{Description of the device}

\begin{figure}
\includegraphics[width=0.5\textwidth]{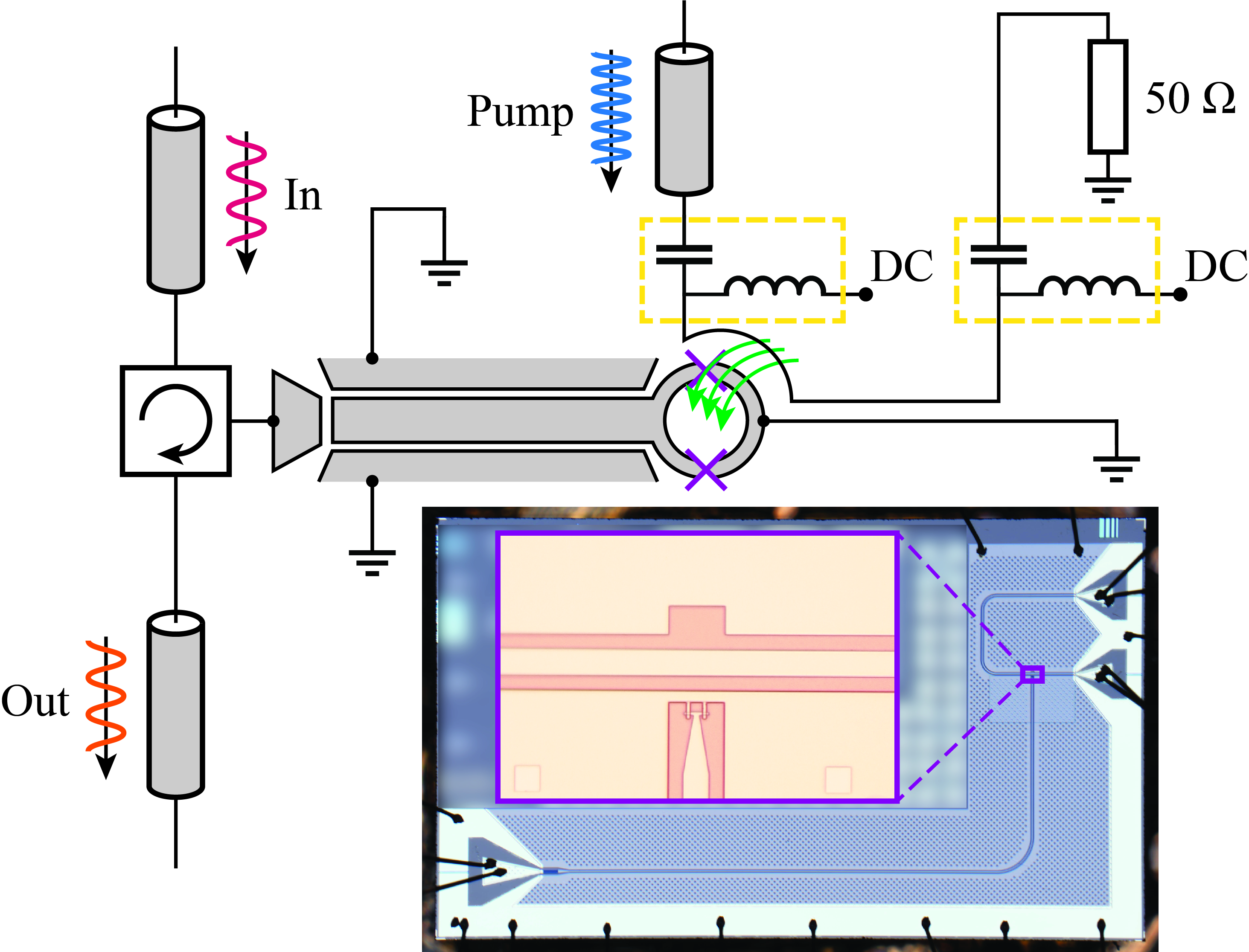}
\caption{{\bf Schematic of the sample and the low-temperature configuration of the experimental setup.} The flux-modulated K-JPA is pumped at twice the frequency of the probe signal. The probe signal propagates through the circulator into the parametric cavity. Bias-tees allow to combine the RF-pump and the DC-control of the detuning. Inset: optical image of the sample with a closeup of the SQUID and feed line.}
	\label{fig:schematic}
\end{figure}

The K-JPA device consists of a $\lambda/4$ coplanar waveguide $50\,\Omega$ resonator terminated with a SQUID loop, see Fig.~\ref{fig:schematic}. The inset shows the SQUID loop consisting of two JJs and the flux line used for pumping and flux biasing. The tunnel junctions and wiring layers were fabricated at VTT Technical Research Centre of Finland by Nb and Al sputtering on a high-resistivity silicon substrate \cite{Gronberg_2017}. The value of the critical current for the Nb/Al-Al$_2$O$_3$/Nb $1\times1\,\unit{\micro\meter^2}$  JJs is $I_\mathrm{c}\approx8\,\unit{\micro\ampere}$.
The measurements were done at 20 mK in a BlueFors dry dilution cryostat, where the device was mounted in  a sample holder and protected from external magnetic noise  with a Cryoperm shield, see Appendix A. 
The frequency of the resonator can be tuned by applying a magnetic field, $\omega_{0}(\Phi_\mathrm{DC})\approx{\omega_{\lambda/4}}/[1+\chi(\Phi_\mathrm{DC})]$ \cite{Krantz_2013,pogorzalekFluxdrivenJosephsonParametric2017},  where $\omega_{\lambda/4}/(2\pi)=6.179\,\unit{\GHz}$ is the bare resonance frequency in the absence of the SQUID, $\Phi_\mathrm{DC}$ is the applied magnetic flux,  and $\chi(\Phi_\mathrm{DC})$ $=L_\text{S}(\Phi_{\mathrm{DC}})/L_\text{cav}$ is the inductive participation ratio of the Josephson inductance. The range of frequencies span from a maximum of 6.117\,GHz at zero field to below 5.5\,GHz.  At the chosen operational point we have $\Phi/\Phi_0\approx 0.362$, where $\Phi_{0}=h/(2e)$ is the magnetic flux quantum, resulting in $\chi_0=0.0227$ and $\omega_{0}/(2\pi)=6.042\,\unit{\GHz}$. The K-JPA is pumped with a microwave tone applied via inductive coupling to the SQUID at a frequency $\omega_\text{P}$ close to $2\omega_0$, resulting in three-wave mixing \cite{yamamoto2016parametric, Elo2019}. The DC flux bias and the RF pump share a common on-chip flux line, and the signals are combined by an external bias-tee. The inner conductor of the line is separated from the common ground of the sample and cryostat (see Fig. \ref{fig:schematic}), in order to avoid additional heating by ground currents. The input signals are applied via a circulator with a $4-8$\,GHz frequency band. The pulses are phase locked and the timings are synchronized by external triggers with high accuracy.
The readout is done by using a standard heterodyne scheme, employing a local oscillator tone at $f_{\rm LO}=6.028\,\unit{\GHz}$ which is modulated at an intermediate frequency $f_{\rm IF}=14\,\unit{\MHz}$.  
                                                                                                                                                                                                                                                                                                                                                                                                                                                                                                                                                                                                                                                                                                                                                                                
In a frame rotating at half the pump frequency and in the rotating wave approximation, the Hamiltonian of the system reads
\begin{equation}
	H_{\rm sys}^{\rm (RWA)} = \frac{\hbar \Delta}{2}(\mathcal{P}^2 + \mathcal{Q}^2)
	+ \frac{\hbar |\alpha |}{2} (\mathcal{P}\mathcal{Q} +
	\mathcal{Q}\mathcal{P} ) + \frac{3\hbar K}{2} (\mathcal{P}^2 + \mathcal{Q}^2)^2,\label{eq:HRWA}
\end{equation}
where $\mathcal{Q} = (a + a^{\dag})/\sqrt{2}$, $\mathcal{P} = -i(a - a^{\dag})/\sqrt{2}$, are the dimensionless quadratures of the cavity field 
$a$, $\alpha$ is the pump amplitude, and $K$ is the Kerr nonlinearity. The detuning $\Delta = \omega_{0}-\omega_{\text{P}}/2$ is defined with respect to half the pump frequency $\omega_\text{P}$. The coupling constant of the probe signal into the device is denoted by $\kappa$ and the internal dissipation is characterized by a rate $\gamma$. From characterization measurements of the spectrum and gain coefficient we extract $\kappa /(2\pi) = 4.44\,\unit{\MHz}$,  $\gamma/(2\pi)=2.30\,\unit{\MHz}$, and we get  $K/(2\pi) \approx -0.2\,\unit{\kHz}$ by fitting with a simulation of the quantum Langevin equation and by calculations based on circuit parameters.
The next-order term is sextic, and from circuit analysis we obtain a value of a few orders less than the Kerr constant, $\Lambda\sim10^{-6} |K|$, therefore we will neglect it. Overall, the device is remarkably robust under cryogenic cycling and aging: over a period of three years, we have not observed significant changes in bandwidth, gain, or flux-dependency of the spectrum.

For certain values of the parameters $(\Delta , \alpha)$ the system with the Hamiltonian Eq. (\ref{eq:HRWA}) may become unstable, resulting in phase transitions to oscillatory states, as described in the next section.

\section{Phase transition diagram}

The existence of phase transitions between different phases (or states) of a parametrically pumped nonlinear oscillator has been predicted theoretically \cite{lin2015critical, marthaler2007quantum, peano2012sharp, wustmannParametricResonanceTunable2013a, dykmanFluctuatingNonlinearOscillators2012}. 
In order to analyze the phase transitions and map the full phase diagram, we capture the quadrature voltage amplitudes $X$ and $Y$ by heterodyning the amplified output of the device, which we express hereafter in arbitrary units.  For every value of the pump amplitude $\alpha$ and the flux bias $\Phi/\Phi_0$ we acquire the raw $X,Y$ data over a relatively long time duration of $0.1$\,s. We sweep the pump strength $\alpha/(\kappa+\gamma)$ from 0.3 to 0.75 and the flux bias $\Phi/\Phi_0$ in the range from 0.352 to 0.37 (corresponding to detuning values $\Delta/2\pi$ from $-6\,\unit{\MHz}$ to $6\,\unit{\MHz}$). Note that increasing the flux causes a decrease of the  resonator frequency $\omega_0$ and therefore a decrease of detuning, while the pump/readout frequency is fixed in experiment. Since the absolute phase of the complex signal $X+iY$ can be chosen arbitrary, we rotate it for convenient visualization such that the stretching and displacement are oriented along the $X$ axis. We calculate the first and the second order statistical moments of the $X,Y$ distributions and characterize the states in steady-state measurements with respect to the average values $\left<X\right>,\left<Y\right>$ and the variances $\left<X^2\right>-\left<X\right>^2$, $\left<Y^2\right>-\left<Y\right>^2$.
In Table \ref{states} we show the resulting classification of these states: vacuum state, squeezed vacuum state, unstable oscillation, and coherent-oscillation state (also called "period-2 vibrations" \cite{lin2015critical}).
\begin{table}[h!]
	\begin{tabular}{ |c|c| c| c |c| } 
		\hline \hline
		State & $\left<X\right>$ & $\left<Y\right>$ & $\left<X^2\right>-\left<X\right>^2$ & $\left<Y^2\right>-\left<Y\right>^2$ \\
		\hline \hline
		Vacuum & 0 & 0 & $\left<X_\text{vac}^2\right>$ & $\left<Y_\text{vac}^2\right>$ \\ 
		Squeezed vacuum & 0 & 0 &  $>\left<X_\text{vac}^2\right>$ & $<\left<Y_\text{vac}^2\right>$ \\ 
		Unstable oscillation & 0 & 0 & $\gg\left<X_\text{vac}^2\right>$ & $\approx \left<Y_\text{vac}^2\right>$ \\
		Coherent oscillation & $\gg 0$ & 0 & $\approx \left<X_\text{vac}^2\right>$ & $\approx \left<Y_\text{vac}^2\right>$ \\
		\hline
	\end{tabular}
	\caption{State classification by the mean and variance of quadratures with the vacuum state taken as reference.}
	\label{states}
\end{table}

\begin{figure*}
	\includegraphics[width=1\linewidth]{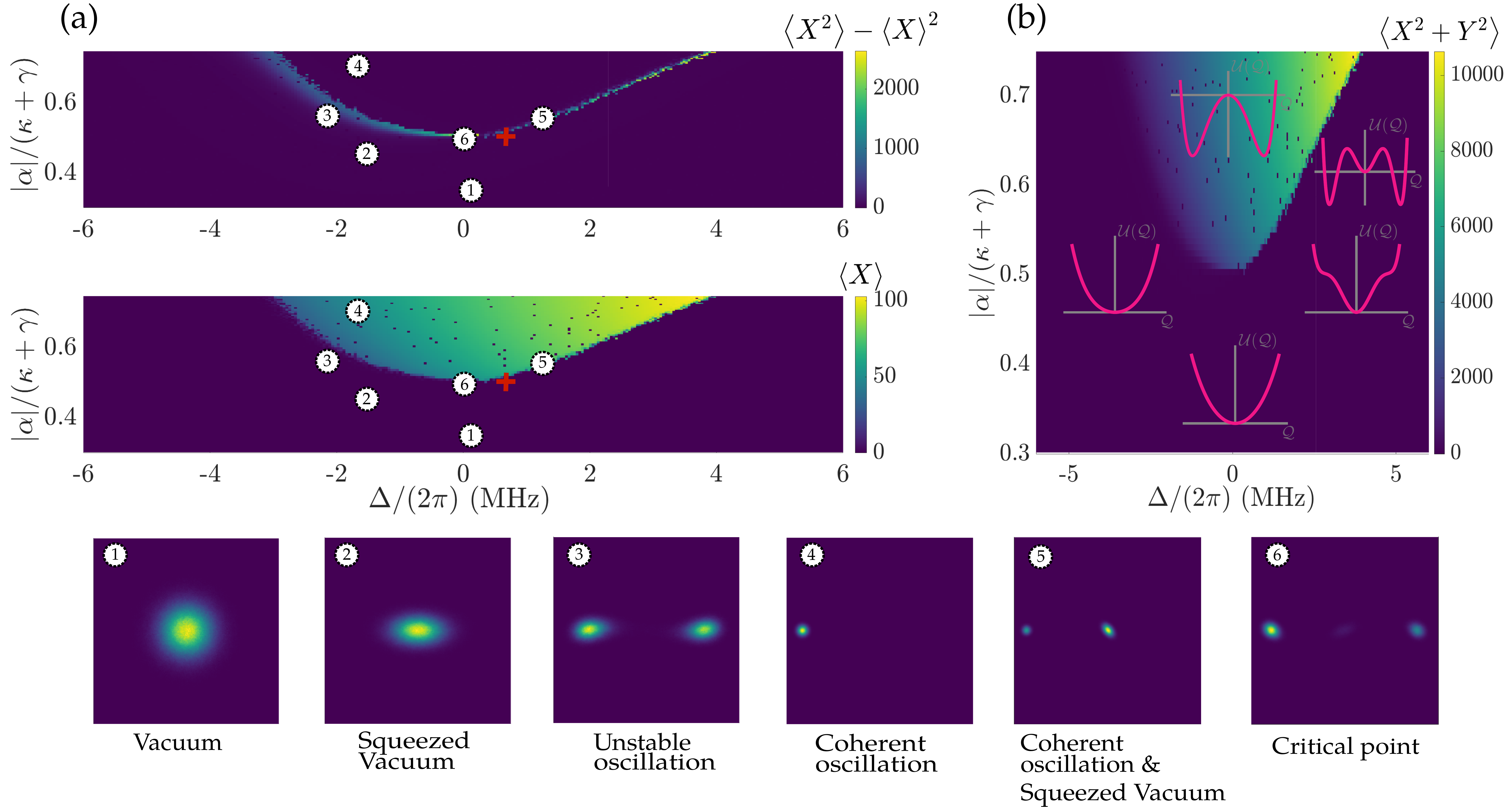}
	\caption{{\bf Experimentally-obtained phase diagram.} 
		The different states of the device can be diagnosed by the statistical properties of the output quadratures $X$ and $Y$, see also Table \ref{states}. (a) The high-contrast region in the plot of the variance of $X$ (a, top) shows the unstable oscillation, while the high-contrast region in the plot of the mean (a, bottom) shows the coherent oscillation. The region below the border is vacuum or squeezed vacuum. 
		At zero frequency detuning one can notice the critical point where the squeezed vacuum, the unstable oscillation, and the coherent oscillatory phase merge. The red crosses mark the operational point for detection. (b) Photon output power dependency over pump amplitude and detuning sweep. The diagrams represent the effective potential $\mathcal{U}(\mathcal{Q})$ as a function of the slow variable $\mathcal{Q}$ in each relevant region.
		The lower plots numbered from 1 to 6 are histograms of the output $X+iY$, and they were collected at the corresponding points in the upper phase diagram. These histograms are obtained by sampling the quadratures over a single long measurement;  they clearly show distinct behaviors of the device under different detunings and pump strengths, indicating different phases.
	 }
	\label{QPTMeanVar}
     \end{figure*}

The phase diagram obtained in this way is presented in Fig. \ref{QPTMeanVar}; similar results were also reported in Ref. \cite{wilsonPhotonGenerationElectromagnetic2010}. The phase diagram agrees very well with the theoretical predictions based on identifying a slow variable $\mathcal{Q}$ and its associated effective potential $\mathcal{U} (\mathcal{Q})$ (see Appendix B and the diagrams in Fig. \ref{QPTMeanVar}(b)). We see that -- since the data at each point is acquired in a long continuous measurement -- the system has sufficient time to explore the effective potential and tends to be partially trapped in the deepest potential wells.

By changing the  pump amplitude and the flux biasing we can control the state of the K-JPA near the critical threshold as follows:
 
 \begin{itemize}
 	\item A second-order transition exists at negative detunings $\Delta<0$. The system changes smoothly from the squeezed vacuum to the region of unstable $0-\pi$ oscillations, see Fig. \ref{QPTMeanVar}(a), point $\circled{3}$.
 	\item A first-order phase transition is present at positive detunings $\Delta>0$ (see for example point $\circled{5}$ in Fig. \ref{QPTMeanVar}(a); here the mean value has a jump $\left<X\right>=0\rightarrow\left<X\right>\gg0$. This transition border is definitely sharp and suitable for defining the operational point for photon detection. 
 	\item A critical point is observed at $\Delta = 0$ and $\alpha/(\kappa + \gamma) = 0.5$ where the two types of phase transitions above merge, see Fig. \ref{QPTMeanVar}(a) point $\circled{6}$.
 \end{itemize}
 
For the detection of weak signals approaching the level of single photons, the first-order transition border at positive detunings $\Delta= \omega_0 -\omega_\text{P}/2$ and critical pump amplitude value $\alpha_{c} (\Delta) = \sqrt{\Delta^2+(\kappa+\gamma)^2/4}$ is preferable owing to its sharpness and reliable discrimination of switching events due to the large average quadrature $\langle X \rangle$. Three local minima exist in the effective potential when approaching from the right of the transition borderline, similar to the Landau theory of first-order transitions. In contrast, the second-order transition appearing at negative detunings is characterized by a smooth change from the zero-amplitude state to states with relatively small positive or negative quadratures.

To observe transitions from squeezed vacuum to coherent oscillation experimentally, we have monitored the in-phase quadrature $X$ for small powers of the incoming probe field, and recorded sudden bursts to large values. Several operational points were probed at different positive detunings; the suitable ones should have a reasonable small switching probability in the absence of probe excitation (dark count rate) and also should present a measurable change in these probabilities when the probe field is turned on. Finally, we decided for an operational point close to the critical point \circled{6}, at a flux $\Phi /\Phi_0 \approx 0.3618$ (positive detuning $\Delta/(2 \pi) = 0.7\,\unit{\MHz}$) and pump amplitude $\alpha/(\kappa+\gamma)\approx 0.51$, marked with a red cross in Fig. \ref{QPTMeanVar}.
 
Next, the detection protocol can be devised as follows. In the central well of the effective potential we prepare the state characterized by zero mean amplitude $\left<X\right>=0$ (squeezed vacuum state).
In the absence of a probe signal, the state of device can change by activation over the barrier due to quantum and thermal fluctuations \cite{lin2015critical,dykman2007critical}; these events are called dark counts. If a microwave probe pulse is applied, the effective potential gets tilted and the activation is easier (see Appendix D).

\section{Photon detection}

We can now demonstrate single-photon detection at the first-order phase transition.
In this case, the perturbation triggering the switching is provided by the absorbed photon.

\begin{figure*}
\includegraphics[width=1\linewidth]{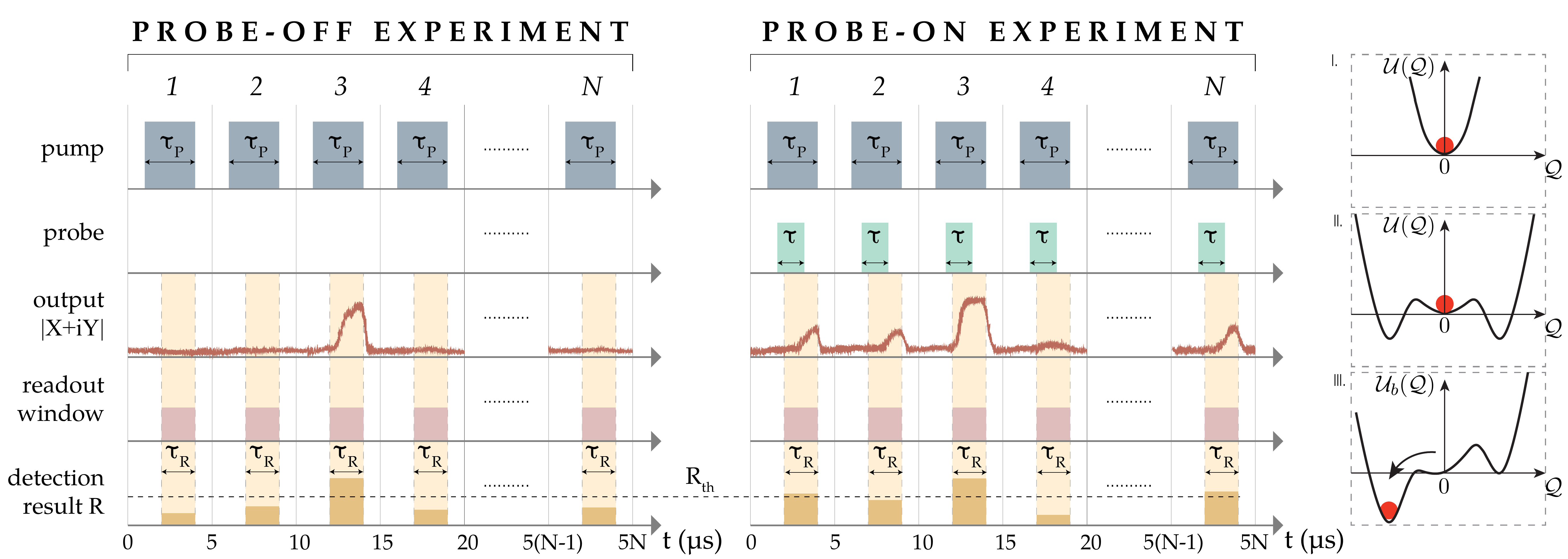}
\caption{{\bf Detection protocol.} Pulse sequence for the probe pulse turned OFF or ON, with the illustrations on the right showing the effective potential representation as a function of the slow quadrature $\mathcal{Q}$ (see Appendix B). Each measurement starts by sending  a pump pulse of duration $\tau_{\rm P}$ and frequency $\omega_\text{P}$, with the system initiated in the ground state (\textit{I}), in order to bring it at the operational point (\textit{II}). After a small 
delay time the probe pulse of duration $\tau$ is applied, resulting in a tilted effective potential $\mathcal{U}_{b}$ and activation out of the central well (\textit{III}). 
The detection session ends by switching off the pump in order to bring the cavity back to the ground state (\textit{I}). The measurement yields a detection result $R$  defined as the time-average of $| X+iY |$ during the readout window width $\tau_{R}$. 
The role of the PROBE-OFF experiments is to measure the dark count probability $p_\texttt{dark}$. In both the PROBE-ON and PROBE-OFF experiments, the decision about positive detection events is taken by comparison of $R$ with a chosen threshold value $R_{\rm th}$.}
\label{waveforms}
\end{figure*}

\begin{figure}
	\includegraphics[ width=1\linewidth]{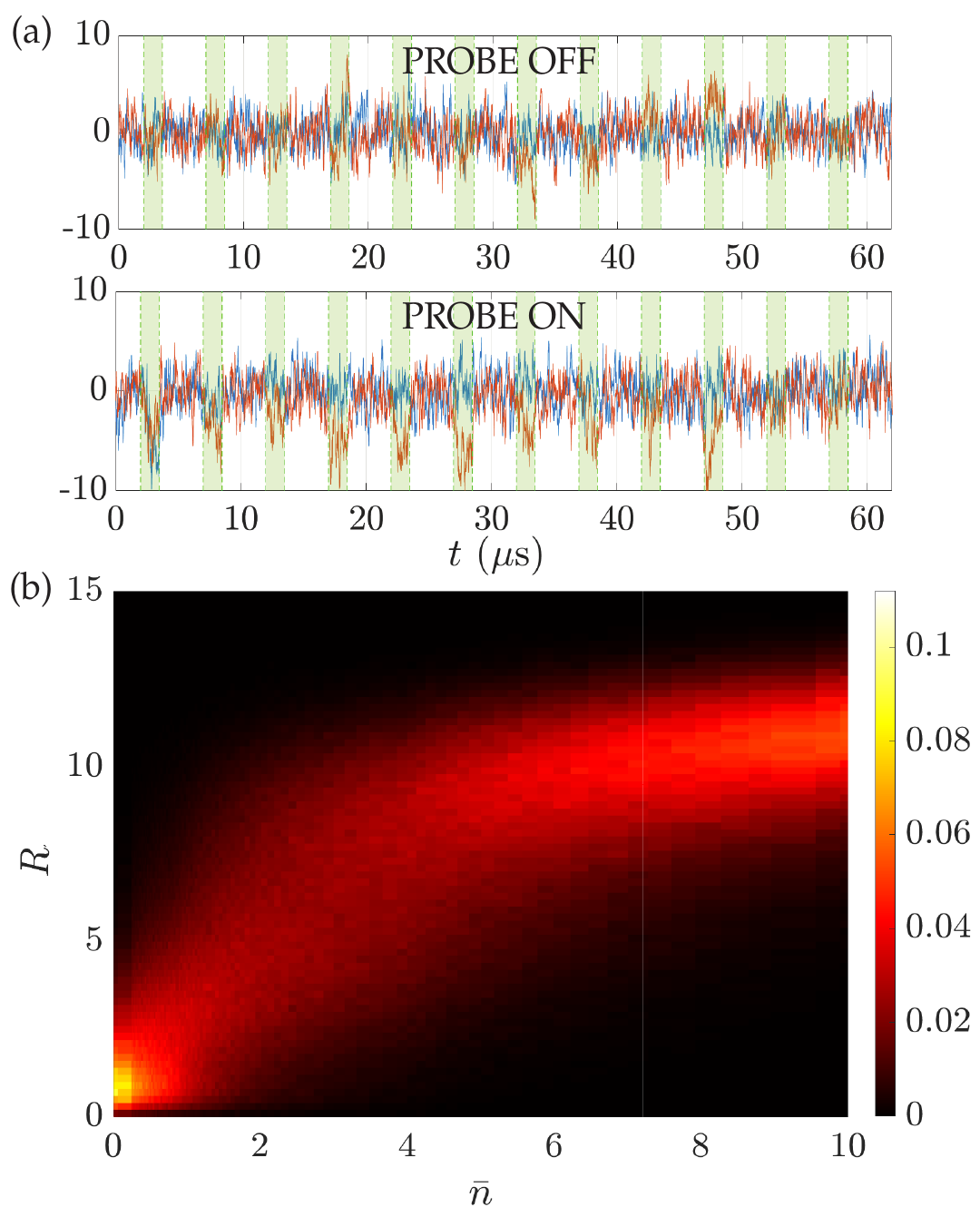}
	\caption{{\bf Quadratures and thresholds.} (a) We present a sample of measured  quadratures $(X,Y)$ (arbitrary units) in time-domain. The green bands denote the readout windows of duration $\tau$. (b) Normalized histograms of detection events yielding the results $R$, recorded at each discretized value of the average number of photons $\bar{n}$. The duration of the pulses was $\tau = 1\,\unit{\micro\second}$ and for each $\bar{n}$ we used a total number $N=2\times10^4$ of pulses.}
	\label{histograms_detection}
\end{figure}

\paragraph{Pulse sequence.}
The protocol uses a specific sequence of pulses for the pump, the probe, and for the readout window, as presented schematically in Fig.	\ref{waveforms}.
The detection procedure starts with the unpumped device in the ground (vacuum) state, after which we bring it at the operational point
($\alpha/(\kappa+\gamma)\approx0.51$, $\Phi/\Phi_0 \approx 0.3618$, $\Delta /(2\pi) = 0.7\,\unit{\MHz}$) by ramping up the pump. 
We send a rectangular-shaped calibrated probe pulse containing $\bar{n}$ photons on average to the input port of the K-JPA and we record the output signal. The presence of the probe pulse results in a tilted effective potential $\mathcal{U}_{b}(\mathcal{Q})$, increasing the probability of escape into one of the outer wells, see Appendix D. During the readout window of duration $\tau_{R} = 1.5\,\unit{\micro\second}$ the recorded trace is averaged to define the detection result $R = (1/\tau_{R} )\int_{0}^{\tau_{R}} dt | X(t)+iY(t) |$. In the final step of the  measurement we switch off the pump, creating the dead time necessary for the cavity to relax back to the ground state. This is needed because the strong oscillations, once established, will not relax by themselves and the  detector will not be able to register another photon.
This dead time is set to $3\,\unit{\micro\second}$, from the falling edge of pump pulse to the rising edge of next pump pulse, sufficient to bring back the parametric cavity to its ground state. The latency time, the dead time, and the readout window value define the minimum time before the next pulse sequence, resulting in a maximum repetition rate of $2\times 10^5\,\mathrm{s}^{-1}$ and a duty cycle of 40\% at $\tau=1\,\unit{\micro\second}$, see Fig. \ref{waveforms}. The calibration of the mean photon number $\bar{n}$ is based on the determination of the gain coefficient in the output line and on a reflection measurement of the attenuation in the input line (see Appendix A). We collect sufficient quadrature data ($N=2\times10^4$ pulses) at every value of $\bar{n}$ -- see Fig. \ref{histograms_detection}(a) -- to obtain reliable statistics and to construct histograms normalized over the number of pulses $N$ in the sequence, as shown in Fig. \ref{histograms_detection}(b).

From these probability distributions one can see that the detection is not photon-number-resolving, due to the overlap of distributions for different $\bar{n} > 1$. Thus, the device is inherently a threshold detector.

\paragraph{SPD basic characterization.}

A threshold detector yields one bit of information corresponding to either of the two states ``switched/click'' or ``non-switched/non-click'', depending  on weather the response of the device is below or above a certain threshold set by the experimentalist \cite{migdallSinglephotonGenerationDetection2013}. Here we denote these two states as $\texttt{0}$ and $\texttt{1\raisebox{-0.5ex}{+}}$, defined by $R < R_{\rm th}$ and $R> R_{\rm th}$ respectively. Here $R_{\rm th}$ is the response threshold. 
Ideally, the detector should react to the presence of a probe coherent field at the input by switching with probability $p_{\texttt{1\raisebox{-0.5ex}{+}}}=1-p_\texttt{0}=1-\exp[-\eta \bar{n}]$, where $\eta$ is the single-photon efficiency (see also Appendix C). In reality, the detector may switch also when no probe signal is present, producing a false positive (dark count); we denote the probability of these events by $p_\texttt{dark}$. As a result, the probabilities of recording $\texttt{0}$ or $\texttt{1\raisebox{-0.5ex}{+}}$ are not the ideal $p_{\texttt{1\raisebox{-0.5ex}{+}}}$ and 
$p_\texttt{0}$. Instead, we have the true-positive probability $P_{\texttt{1\raisebox{-0.5ex}{+}}}$ and its complement
$P_\texttt{0}$ with a structure obtained by the composition rules for probabilities
$P_{\texttt{1\raisebox{-0.5ex}{+}}} \equiv 
p_{\texttt{1\raisebox{-0.5ex}{+}$\lor$dark}} = p_{\texttt{1\raisebox{-0.5ex}{+}}}+p_\texttt{dark}-p_{\texttt{1\raisebox{-0.5ex}{+}}} p_\texttt{dark}$ and $P_\texttt{0} \equiv p_{\texttt{0$\land\neg$dark}}=(1-p_\texttt{dark})p_{\texttt{0}}$, with normalization $P_{\texttt{1\raisebox{-0.5ex}{+}}} + P_\texttt{0} =1$.

From these definitions we can relate the efficiency to the probabilities  $P_{\texttt{1\raisebox{-0.5ex}{+}}}$ and $p_\texttt{dark}$ that are measured directly 
\begin{equation}
	\eta =\frac{1}{\bar{n}}\ln\frac{1-p_\texttt{dark}}{1-P_{\texttt{1\raisebox{-0.5ex}{+}}}}.
	\label{eta}
\end{equation}

For our detector, the efficiency depends weakly on $\bar{n}$; this effect becomes visible at large values of $\bar{n}$. We can include this dependence by the replacement  $\eta \rightarrow \eta_\epsilon (\bar{n})$, with $\eta_\epsilon (\bar{n})$ defined phenomenologically as 
$\eta_\epsilon(\bar{n}) =\eta^{\frac{1}{1+ \epsilon (\bar{n}-1)}}$
where $\eta$ is the efficiency at $\bar{n}\sim 1$ and $\epsilon\ll 1$ is a subunit constant.

\begin{figure*}
	\includegraphics[ width=1\linewidth]{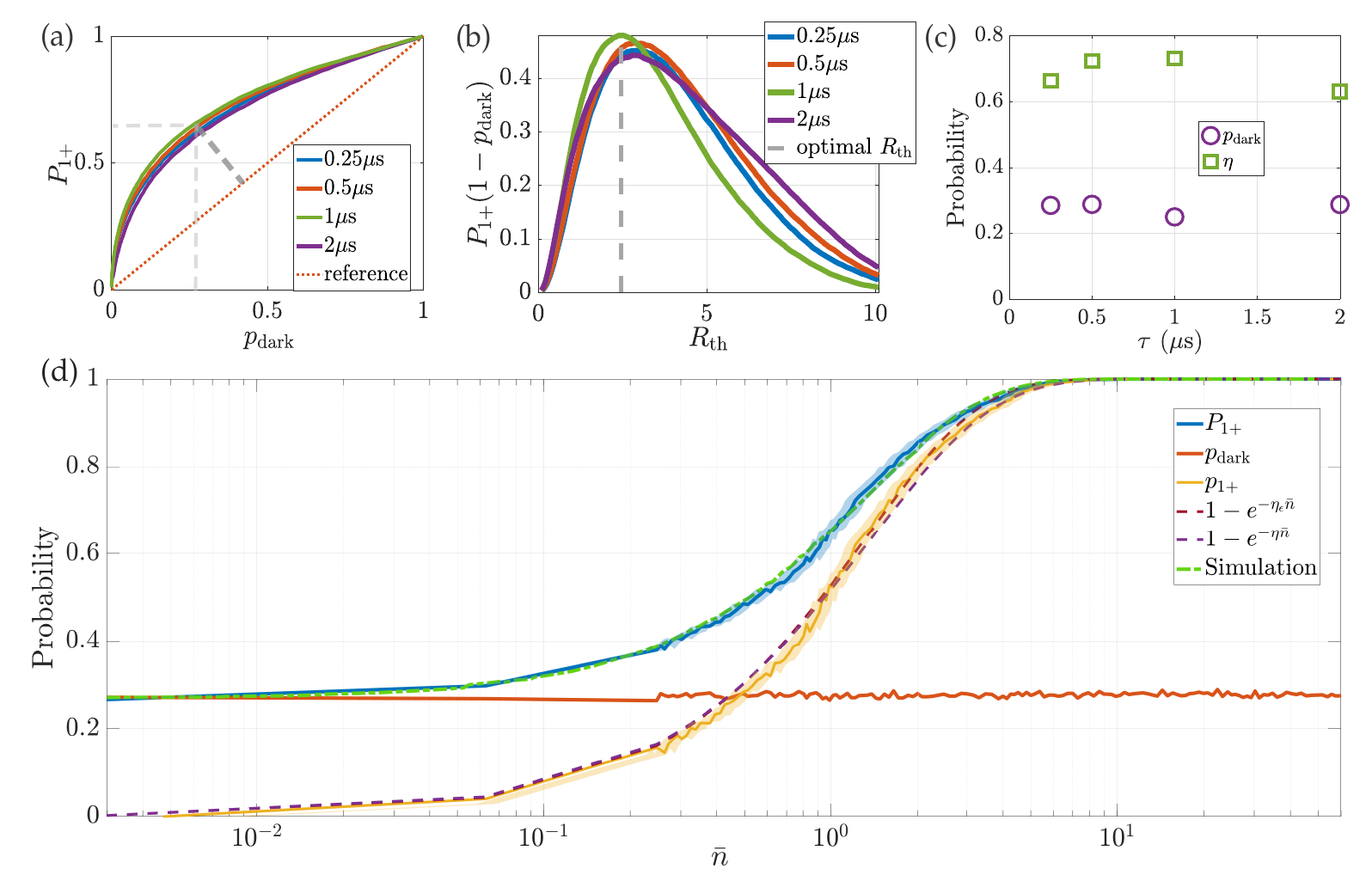}
	\caption{
		{\bf SPD's basic characterization results}. 
		We present: (a) the receiver operating characteristics, with the dashed lines indicating the maximum separation from the zero-efficiency reference, corresponding to the optimal threshold;
		(b) The quantity $P_{\texttt{1\raisebox{-0.5ex}{+}}}(1-p_{\texttt{dark}})$ as a function of threshold; 
		(c) single-photon efficiencies $\eta$ and dark count probabilities  $p_{\texttt{dark}}$ for probe pulses of different duration but with the same average number of photons per pulse $\bar{n} = 1$; 
		(d) measured probabilities $P_{\texttt{1\raisebox{-0.5ex}{+}}}$ and $p_{\texttt{dark}}$
as a function of $\bar{n}$ for probe pulses with the same duration $\tau =1\,\unit{\micro\second}$. 
The probability $p_{\texttt{1\raisebox{-0.5ex}{+}}}$ is obtained from these data as $p_{\texttt{1\raisebox{-0.5ex}{+}}} = (P_{\texttt{1\raisebox{-0.5ex}{+}}} - p_{\texttt{dark}})/(1 - p_{\texttt{dark}})$ and it is fitted with $1 - \exp (-\eta \bar{n})$ and $1 - \exp (-\eta_{\epsilon} \bar{n})$ with $\eta = 0.73$ and $\epsilon =0.1$. The calibration error in the number of photons $\bar{n}$ from one value to another (depicted as yellow or blue semitransparent areas) is estimated as $\pm0.35$\,dB.
}
	\label{main_results}
\end{figure*}

We start by identifying a suitable operational point 
and then optimizing the threshold value $R_{\rm th}$. This is done for different pulse durations $\tau = \lbrace0.25;0.5;1;2\rbrace\,\unit{\micro\second}$, all with $\bar{n} =1$ photons. The protocol \cite{buller2009single,chenMicrowavePhotonCounter2011} consists of the following sequence of measurements (see also Fig. \ref{waveforms}). Firstly, we do not send any probe pulses (PROBE-OFF experiment), obtaining the false positive ratio (dark count probability) $p_{\texttt{dark}}$; secondly, we send $N$ probe pulses containing $\bar{n}$ photons each (PROBE-ON experiment) obtaining the true positive ratio $P_{\texttt{1\raisebox{-0.5ex}{+}}}$. Then we build the receiver operation characteristic with floating threshold value $R_{\rm th}$, see Fig. \ref{main_results}(a). We choose the operational point parameters $(\alpha, \Delta)$ such that the area-under-curve is maximal. The diagonal represents the reference value for a detector with zero efficiency (yielding only dark counts).
Next, we find the optimal value of the threshold by constructing the $P_{\texttt{1\raisebox{-0.5ex}{+}}}(1-p_{\texttt{dark}})$   curves  (see Fig. \ref{main_results}(b). The threshold $R_{\rm th} \approx 2.45$ corresponds to the maximum of the $\tau = 1\,\unit{\micro\second}$ trace and it produces a near-maximum for the others as well. In the representation of Fig. \ref{main_results}(a) , this optimal threshold coincides with the maximum separation of the receiver characteristics from the diagonal reference line.

Next, by using Eq. (\ref{eta}) we can extract the efficiency $\eta$. The results, with the operational point and threshold optimized for each value of $\tau = \lbrace0.25;0.5;1;2\rbrace\,\unit{\micro\second}$ are presented in Fig. \ref{main_results}(c). We can see that the detector can be thus calibrated to photon detection with similar efficiencies $\eta$ and dark count probabilities $p_{\texttt{dark}}$ for pulses of different durations.

A more detailed analysis is shown in Fig. \ref{main_results}(d). Here we present the measured $P_{\texttt{1\raisebox{-0.5ex}{+}}}$ and $p_{\texttt{dark}}$ for a wide span of $\bar{n}$ photons together with numerical simulations using a Fokker-Planck equation describing the dynamics of switching from the effective potential shown in Fig. \ref{waveforms} (see Appendix D). For each value of $\bar{n}$ we measure $P_{\texttt{1\raisebox{-0.5ex}{+}}}$, then, in order to check the stability of the system, we turn off the probe field and measure $p_{\texttt{dark}}$. We extract the ideal probability $p_{\texttt{1\raisebox{-0.5ex}{+}}} = (P_{\texttt{1\raisebox{-0.5ex}{+}}} - p_{\texttt{dark}})/(1 - p_{\texttt{dark}})$ and we fit it with $1 - \exp (-\eta \bar{n})$ and with 
$1 - \exp (-\eta_{\epsilon} \bar{n})$ (here $\epsilon = 0.1$), obtaining $\eta = 0.73$. Furthermore, we observe that at low $\bar{n} \ll 1$ a coherent state   
yields approximately the same probabilities as a state $(1-\bar{n})|0\rangle\langle 0| +\bar{n}|1\rangle \langle 1|$, which  represents a source emitting single photons randomly with probability $\bar{n}$ during the time $\tau$. From the positive operator valued measure (POVM) formalism (see Appendix C) we obtain for both states $p_{\texttt{0}} \vert_{\bar{n}\ll 1}\approx1 -\bar{n}\eta$ and $p_{\texttt{1\raisebox{-0.5ex}{+}}} \vert_{\bar{n}\ll 1}\approx \bar{n}\eta$. This relation can be readily verified from Fig. \ref{main_results}(d), and it has the meaning that the detection probability is the product between the emission probability and the single-photon detection efficiency $\eta$.  In other words, from
Fig. \ref{main_results}(d) we can see that the dark count probability is the intercept of $P_{\texttt{1\raisebox{-0.5ex}{+}}}$ with the vertical axis corresponding to the asymptotic $\bar{n}=0$, while the efficiency is the slope of $p_{\texttt{1\raisebox{-0.5ex}{+}}}$ at the same point,
$\eta = (dp_{\texttt{1\raisebox{-0.5ex}{+}}}/d \bar{n})\vert_{\bar{n}=0}$.

\begin{figure}
	\includegraphics[width=1\linewidth]{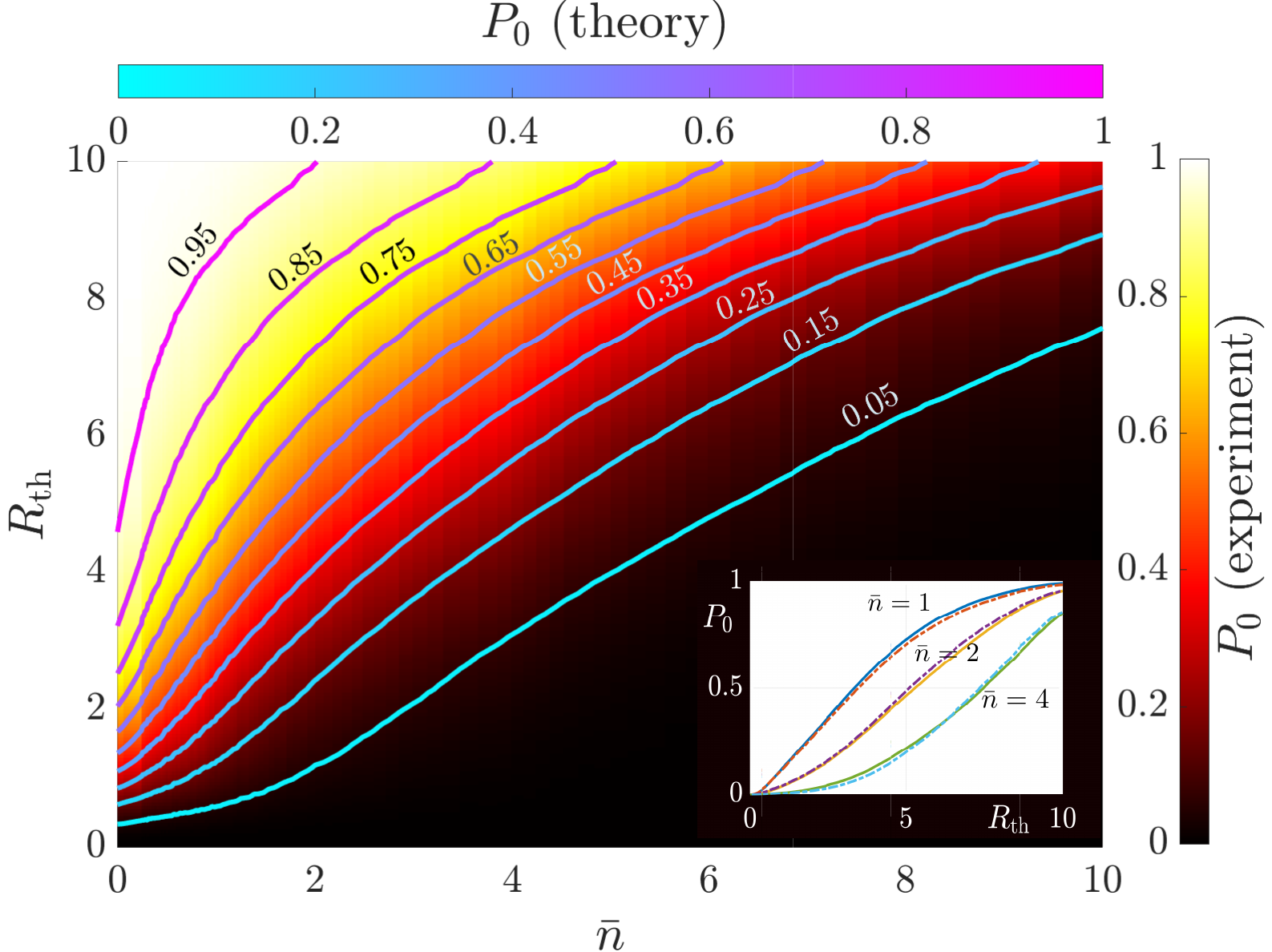}
	\caption{Surface and contour plots demonstrating the Poisson statistics of a probe signal containing $\bar{n}$ photons. We present the overall probability to detect zero photons $P_\texttt{0}$ as a function of  $\bar{n}$ in pulse and threshold value. The data shown as surface plot is the experimental $P_\texttt{0}$, whereas the contour plot is the reconstructed
	$P_\texttt{0}=[\exp (-\eta_{\epsilon}\bar{n})](1-p_\texttt{dark})$ using the values $p_\texttt{dark}$ and $\eta_{\epsilon}$ obtained experimentally for each value of the threshold $R_{\rm th}$. The inset shows three cross-sections along $R_{\rm th}$ at $\bar{n}=1,2,4$. The experimental results are represented with solid lines, and the theory is shown with dash-dot lines. }
	\label{p0_Poisson}
\end{figure}

\paragraph{Noise equivalent power.}

Seen as a power sensor, the sensitivity to the microwave field present at the input can characterized by the power responsivity (susceptibility) at low photon numbers
\begin{equation}
	\left.\frac{dP_{\texttt{1\raisebox{-0.5ex}{+}}}}{
		d\mathbb{P}}\right|_{\bar{n}=0}  =
	\frac{\eta_\epsilon (\bar{n}=0)\tau}{\hbar \omega}(1 -p_{\texttt{dark}}).
\end{equation}
Here $\mathbb{P} = \hbar \omega \bar{n}/\tau$ is the input power in a pulse of duration $\tau$ and containing $\bar{n}$ photons. We obtain a power responsivity of $1.3 \times 10^{17}~\mathrm{W}^{-1}$.

The noise equivalent power (NEP) (see also Appendix C) can be obtained from \cite{Komiyama5475208, Cleland1992, astafievSinglePhotonDetectionQuantum2003}
\begin{equation}
	\mathrm{NEP}_{\epsilon} = \frac{\sqrt{2 \Gamma_{\texttt{dark}}}}{\eta_\epsilon (\bar{n}=0)} \hbar \omega.
\end{equation}
For $\Gamma_{\texttt{dark}} =
0.167\,\unit{\MHz}$ and $\omega/(2\pi ) = 6.042\,\unit{\GHz}$, we get $\mathrm{NEP}_{\epsilon} = \frac{1}{\eta_{\epsilon}(\bar{n}=0)} \times 2.312 \times 10^{-21}~\mathrm{W}/\sqrt{\mathrm{Hz}} = 3.288 \times 10^{-21}\,\mathrm{W}/\sqrt{\mathrm{Hz}}$.

\paragraph{Observation of Poissonian statistics.}

In general, the effect of a subunit efficiency can be understood as a reduction of the number of detected particles by a factor of $\eta$ (see Appendix C). This can be modeled by the use of an attenuator with transmission coefficient $\eta$, which reduces the  average number of particles, followed by detection with unit efficiency \cite{silberhorn2007detecting,All_aume_2004,rossi2004photonstat}.
 Moreover, it is known that the Poissonian character of a coherent state is preserved under attenuation \cite{loudon2000quantum}, as it is apparent from the formulas $p_{\texttt{1\raisebox{-0.5ex}{+}}} = 1 - \exp(-\eta \bar{n})$ and $p_{\texttt{0}}=\exp(-\eta \bar{n})$. Although apparently innocuous, it is not at all guaranteed that this ideal-detector exponential dependence reproduces correctly the response of the real detector, even when corrected for the dark counts.

Here we demonstrate the Poissonian statistics of an input coherent field over a wide range of values of $\bar{n}$ and thresholds $R_{\rm th}$ by comparing the experimentally-measured $P_{\texttt{0}}$ with the values obtained by factoring the measured probability of no dark events and the ideal-detector  no-event probability $\exp(-\eta_\epsilon \bar{n})$, in other words  $P_{\texttt{0}} = [\exp(-\eta_\epsilon \bar{n})](1 - p_\texttt{dark})$.  Here $\eta_\epsilon$ is extracted at each value of the threshold from the measured probability at just one point, $\bar{n}=1$, and $\epsilon=0.1$. Fig. \ref{p0_Poisson} shows this comparison for the case $\tau=1\,\unit{\micro\second}$ over the entire range $\bar{n} \in \{0, 10\}$. In view of the high degree of agreement obtained, we conclude that the results verify convincingly the Poissonian statistics of the incoming photons.

\section{Conclusions}

We have realized a detector of microwave photons by employing the sensitivity to perturbations of a Kerr Josephson parametric amplifier operated near the critical threshold of a first-order phase transition.
We have mapped the full phase diagram and showed that the system is described by an effective potential in the slow quadrature. The operational point is chosen such that the system is in a metastable state near a first-order transition in the proximity of the critical point, and consequently even single-photon microwave pulses
can trigger the system to switch from the squeezed state to an oscillatory state. These oscillations are further amplified to values above the thermal noise and eventually detected using a homodyne scheme at room temperature. Our results include measurement and optimization of threshold, and we further extract the efficiency and the dark count rate. We also show how this detector can be used to observe the Poissonian statistics of photons in a coherent probe field. Our results demonstrate that the effective potential can be controlled and the phase transition can be harnessed to producing a useful device.
Compared to detectors based on switching in current-biased Josephson junctions \cite{chenMicrowavePhotonCounter2011,poudel2012quantum}, our device has a relatively fast repetition rate.  This is limited only by the decay time $2/(\kappa + \gamma) \approx 50$ ns of the resonator . In contrast, Josephson junction detectors switch to the quasiparticle branch of the current-voltage characteristics, and in consequence they are fundamentally constrained by the quasiparticle equilibration time during retrapping, which is several tens of microseconds \cite{Zgirski2019}. In our detector even in the oscillatory state the voltage drops are much smaller than the superconducting gap.
For future experiments, the dark count rate may be further lowered by reducing  technical noise and by increasing the sharpness of first order-transition. This can be done by engineering the effective potential for example by using  SNAIL (Superconducting Nonlinear Asymmetric Inductive
eLement) structures \cite{frattini3waveMixingJosephson2017} to achieve a smaller Kerr coefficient. Our results open the way to using criticality for sensitive detection of electromagnetic fields in superconducting-circuit experimental platforms.

\section*{Data availability}
Data that supporting the findings of this study are
available from authors upon reasonable request. 

\section*{Code availability}
The codes used for this work are available at github.

 \section*{Acknowledgments}

We are grateful to Mark Dykman for helpful clarifications of previous theoretical results, to Visa Vesterinen for reading and commenting the manuscript, as well as to Simone Felicetti, Roberto Di Candia, and Fabrizio Minganti for stimulating  discussions. We acknowledge the team at VTT Technical Research Centre of Finland Ltd for fabricating the device and providing characterization data: Joonas Govenius, Leif Grönberg, Robab Najafi Jabdaraghi, Janne Lehtinen, Mika Prunnila, and Visa Vesterinen. We thank 
Ilari Lilja for contributions to sample design and help with the cryogenic setup.
This project has received funding from the European Union's Horizon 2020 research and innovation programme under grant agreement no. 862644 (FET-Open project QUARTET). We also acknowledge financial support from the Finnish Center of Excellence in Quantum Technology QTF (Projects No. 312295, No. 312296, No. 336810, No. 336813, and No. 352925). We are also grateful for support from Saab, under a research collaboration agreement with Aalto University. This work used the experimental facilities of the Low Temperature Laboratory and Micronova of the OtaNano research infrastructure.

\section*{Author Contributions}

KP and GSP conceived the main idea of the experiment. KP performed the measurements and did the data analysis, while PH was in charge of the cryogenic setup. The theoretical models and numerical simulations were developed by GSP, with additional contributions from JW. The sample was designed by MP and PH. KP, JW, and GSP wrote the manuscript. GSP supervised the project. All authors discussed the results and the final manuscript.

\section*{COMPETING INTERESTS}
The authors declare no competing interests.

\vspace{2cm}

\section*{Appendix A: Experimental setup}

\subsection*{\bf System parameters and Hamiltonian} 

To model our system, which is essentially a quarter-wavelength coplanar waveguide cavity terminated with a SQUID,  we adopt the approach used in Refs. \cite{linJosephsonParametricPhaselocked2014,wallquistSelectiveCouplingSuperconducting2006,eichlerControllingDynamicRange2014,wustmannParametricResonanceTunable2013a}. 
The full classical Lagrangian is

\begin{equation}
	\begin{aligned}
		\mathcal{L}_\text{sys}&=\mathcal{L}_\text{cav}+\mathcal{L}_\text{SQUID},
	\end{aligned}
\end{equation}
where
\begin{equation}
	\left\{
	\begin{aligned}
		\mathcal{L}_\text{cav}&=\sum_{j=1}^{N}{\Delta x\left(\frac{\hbar}{2e}\right)^2\left[\frac{C_l}{2}{\dot{\phi}_j}^2-\frac{1}{2L_l}\frac{\left(\phi_{j}-\phi_{j-1}\right)^2}{(\Delta x)^2}\right]},\\
		\mathcal{L}_\text{SQUID}&=\sum_{j=1,2}^{}{\left[\left(\frac{\hbar}{2e}\right)^2\frac{C_{\text{J},j}}{2}{\dot{\phi}_{\text{J},j}}^2+E_{\text{J},j}\cos{\phi_{\text{J},j}}\right]}.
	\end{aligned}
	\right.
\end{equation}
Here $C_l$ and $L_l$ are the capacitance respectively inductance per unit length of the coplanar waveguide of total length $d$ - therefore the total capacitance is $C_{\rm cav}=C_{l} d$ and the total inductance is $L_{\rm cav}= L_{l} d$. 
Since the two JJs in the SQUID are approximately identical, the SQUID Lagrangian $\mathcal{L}_\text{SQUID}$ simplifies to
\begin{equation}
	\begin{aligned}
		\mathcal{L}_\text{SQUID}
		&=\left(\frac{\hbar}{2e}\right)^2\frac{C_\text{S}}{2}\left[{\dot{\phi}_\text{S}}^2+\left(\pi\frac{\dot{\Phi}}{\Phi_0}\right)^2\right]+E_{\text{S}}\left(\Phi\right)\cos{\phi_\text{S}}\\
		&\approx\left(\frac{\hbar}{2e}\right)^2\frac{C_\text{S}}{2}{\dot{\phi}_\text{S}}^2+E_{\text{S}}\left(\Phi\right)\cos{\phi_\text{S}}.
	\end{aligned}
\end{equation}
where we denote $C_\text{S}=2C_\text{J}$, $\phi_\text{S}=\left(\phi_{\text{J},1}+\phi_{\text{J},2}\right)/2$, and $E_\text{S}(\Phi)=2E_\text{J}\cos(\pi\Phi/\Phi_0)$, and the term containing $\dot{\Phi}$ can be dropped if the flux variation is adiabatic \cite{wallquistSelectiveCouplingSuperconducting2006}.

Noting that $\phi\left(d,t\right)=\phi_N\left(t\right)=\phi_\text{S}\left(t\right)$, the Euler-Lagrange equation at the boundary ($x=d$) gives
\begin{equation}
	\label{boundary_condition}
	\begin{aligned}
		& \left(\frac{\hbar}{2e}\right)^2C_\text{S}{\ddot{\phi}\left(d,t\right)}
		+\left(\frac{\hbar}{2e}\right)^2\frac{1}{L_l}\phi'\left(d,t\right)
		+E_\text{S}\left(\Phi\right)\sin\left[\phi\left(d,t\right)\right]
		\\ & =0.
	\end{aligned}
\end{equation}
In the continuum limit, the Lagrangian of our system becomes
\begin{equation}
	\label{L_sys}
	\begin{aligned}
		\mathcal{L}_\text{sys}&=
		\left(\frac{\hbar}{2e}\right)^2\int_{0}^{d}{\left\{\frac{C_l}{2}{\left[\dot{\phi}\left(x,t\right)\right]}^2-\frac{1}{2L_l}\left[\phi'\left(x,t\right)\right]^2\right\}dx} \\
		&
		+\left(\frac{\hbar}{2e}\right)^2\frac{C_\text{S}}{2}{\dot{\phi}_\text{S}}^2+E_{\text{S}}\left(\Phi\right)\cos{\phi_\text{S}}.
	\end{aligned}
\end{equation}

In the bulk of the waveguide, using the Euler-Lagrange equation for the Lagrangian density in Eq. (\ref{L_sys}), $\partial_t(\partial\mathscr{L}/\partial\dot{\phi})+\partial_x(\partial\mathscr{L}/\partial\phi')-\partial\mathscr{L}/\partial\phi=0$, we get the wave equation
\begin{equation*}
	\begin{aligned}
		\ddot{\phi}(x,t)-v^2\phi''(x,t)=0,
	\end{aligned}
\end{equation*}
which takes us to the ansatz $\phi(x,t)=\phi\sin(kvt)\cos(kx)$, where $v=1/\sqrt{L_lC_l}=d/\sqrt{L_\text{cav}C_\text{cav}}$ is the phase velocity for microwave propagating inside the cavity.

For our purpose, it is convenient to decompose the time-dependent parameter $\Phi$ into a static bias part plus a small flux pumping part, i.e., $\Phi=\Phi_\text{DC}+\Phi_\text{P}\left(t\right)$, which means
\begin{equation}
	\label{decomp}
	\begin{aligned}
		\pi\frac{\Phi}{\Phi_0}&=\pi\frac{\Phi_\text{DC}}{\Phi_0}+\pi\frac{\Phi_\text{P}\left(t\right)}{\Phi_0}.	
	\end{aligned}
\end{equation}

Under DC biasing we can make a linear approximation, i.e., $\sin\left[\phi\left(d,t\right)\right]\approx\phi\left(d,t\right)$, for the boundary condition Eq. (\ref{boundary_condition}) and plug the ansatz into it. We obtain a constraint on the eigenmodes $kd$:
\begin{equation}
	\label{constraint}
	\begin{aligned}
		kd\tan\left(kd\right)=\frac{E_\text{S}\left(\Phi_\text{DC}\right)}{E_\text{L,cav}}-\frac{C_\text{S}}{C_\text{cav}}\left(kd\right)^2,
	\end{aligned}
\end{equation}
where $E_\text{L,cav}=\left(\frac{\hbar}{2e}\right)^2\left/L_\text{cav}\right.$ and $E_\text{S}(\Phi_\text{DC})=2E_\text{J}\cos \left(\pi\Phi_\text{DC}/\Phi_0\right)$.
For the fundamental mode $k_0$, the capacitive term on the right hand side  $-C_\text{S}\left(k_{0}d\right)^2/C_\text{cav}$, can be neglected \cite{wustmannParametricResonanceTunable2013a}. Taking a series expansion for LHS, we can get the approximate solution
\begin{equation}
	\label{mode}
	\begin{aligned}
		k_0d\approx\frac{\pi/2}{1+\chi(\Phi_{\text{DC}})}\qquad 
	\end{aligned}
\end{equation}
with the inductive participation ratio $\chi(\Phi_{\text{DC}})$ defined as
\begin{equation}
	\label{mode}
	\begin{aligned}
	\chi(\Phi_{\text{DC}})\equiv\frac{E_\text{L,cav}}{E_\text{S}\left(\Phi_\text{DC}\right)}=\frac{L_\text{S}(\Phi_{\text{DC}})}{L_\text{cav}}\ll1,
	\end{aligned}
\end{equation}
where $L_\text{S}(\Phi_{\text{DC}})\equiv\frac{\Phi_{0}}{4\pi I_\text{c}\cos\left(\pi\Phi_{\text{DC}}/\Phi_{0}\right)}$.

Let us return to the Lagrangian and only consider a single mode, $\phi(x,t)=\phi_k(t)\cos(kx)$. We therefore have $\phi(d,t) = \phi_\text{S}(t) = \phi_{k} \cos (kd)$. By employing the linearized DC constraint condition Eq. (\ref{constraint}), the Lagrangian $\mathcal{L}_\text{sys}$ takes the form of a nonlinear parametric oscillator
\begin{equation}
	\label{L_nlo}
	\begin{aligned}
		\mathcal{L}_\text{sys}&\approx
		\left(\frac{\hbar}{2e}\right)^2\left[\frac{C_k}{2}{{\dot{\phi}_k}^2\left(t\right)}-\frac{1}{2L_k}{\phi}^2_k\left(t\right)\right] \\
	&	+E_{\text{S}}\left(\Phi\right)\cos{\phi_\text{S}}
		+\frac{1}{2}E_{\text{S}}\left(\Phi_\text{DC}	
		\right)\phi_\text{S}^2 ,
	\end{aligned}
\end{equation}
where
\begin{equation}
	\left\{
	\begin{aligned}
		L_{k}^{-1}&=\frac{\left(kd\right)^2}{2L_\text{cav}}M_k,\\
		C_{k}&=\frac{C_\text{cav}}{2}M_k,
	\end{aligned}		\right.
\end{equation}
with
\begin{equation}
	\qquad\text{with}~
	M_{k}\equiv\left[1+\frac{\sin\left(2kd\right)}{2kd}+\frac{2C_\text{S}}{C_\text{cav}}\cos^2\left(kd\right)\right].
\end{equation}

Also, for the fundamental mode $k_0$, we have $M_0\approx1+\chi(\Phi_{\text{DC}})$.
After introducing the conjugate momentum $n_k=(1/\hbar)\partial{\mathcal{L}_\text{sys}}/\partial\dot{\phi}_k=[\hbar C_k/(2e)^2]\dot{\phi}_k$, by expanding $\cos{\phi_\text{S}}$ to the sixth order and taking advantage of small pump and field amplitudes, i.e., $\pi \Phi_\text{P} /\Phi_{0} \ll 1$ and $\phi_\text{S}\ll1$, we arrive at the Hamiltonian for the parametric oscillator
\begin{equation}
	\begin{aligned}
		\mathcal{H}_\text{sys}&\approx
		\frac{\left(2e\right)^2}{2C_k}n^2_k+\frac{\hbar^2}{2\left(2e\right)^2L_k}{\phi}^2_k \\
		&-\frac{\pi}{2 \Phi_{0}}\Phi_\text{P}(t)\tan\left(\frac{\pi \Phi_\text{DC}}{\Phi_{0}} \right) E_{\text{S}}\left(\Phi_\text{DC}\right)\cos^2\left(kd\right)\phi_k^2 \\
		&\quad
		-\frac{E_{\text{S}}\left(\Phi_\text{DC}\right)}{24}\cos^4\left(kd\right){\phi_k^4}
		+\frac{E_{\text{S}}\left(\Phi_\text{DC}\right)}{720}\cos^6\left(kd\right){\phi_k^6}.
	\end{aligned}
\end{equation}

The quantized Hamiltonian can be rewritten in terms of bosonic creation and annihilation operators (with commutation relations $\left[\phi_k,n_k\right]=i$ and $\left[a,a^\dagger\right]=1$), defined by
\begin{equation}
	\begin{aligned}
		\phi_k &=\frac{2\pi}{\Phi_0}\sqrt{\frac{\hbar}{2\omega_k C_k}}\left(a+a^\dagger\right)
		\equiv\phi_{\text{zpf},k}\left(a+a^\dagger\right), \\	
		n_k &=i\frac{\Phi_{0}}{2\pi}\sqrt{\frac{\omega_k C_k}{2\hbar}}\left(a^\dagger-a\right)
		\equiv in_{\text{zpf},k}\left(a^\dagger-a\right),
	\end{aligned}
\end{equation}
where $\omega_k=1/\sqrt{L_k C_k}=kd/\sqrt{L_\text{cav}C_\text{cav}}$. 
Anticipating the pump frequency $\omega_\text{P}\approx2\omega_k$, we may write $\Phi_\text{P}(t)=\Phi_\text{P} e^{-i\omega_\text{P}t}+\Phi_\text{P} ^*e^{i\omega_\text{P}t}$ and obtain
\begin{equation}
	\begin{aligned}
		{H}_\text{sys}\left(t\right) &\approx
		\hbar\omega_ka^\dagger a
		+\frac{\hbar}{2}\left(\alpha_k e^{-i\omega_\text{P} t} + \alpha^{*}_k e^{i\omega_\text{P} t}\right)\left(a+a^\dagger\right)^2 \\
		& +\hbar K_k\left(a+a^\dagger\right)^4
		+\hbar \Lambda_k\left(a+a^\dagger\right)^6,
	\end{aligned}
\end{equation}
in which
\begin{equation}
	\label{nonlinear_coeff}
	\left\{
	\begin{aligned}
		\alpha_k&=-\frac{\pi \Phi_\text{P}E_{\text{S}}\left(\Phi_\text{DC}\right)
		}{\hbar \Phi_{0}}\tan\left(\frac{\pi \Phi_\text{DC}}{\Phi_{0}} \right)\cos^2\left(kd\right)\phi_{\text{zpf},k}^2,\\
		K_k&=-\frac{E_{\text{S}}\left(\Phi_\text{DC}\right)}{24\hbar}\cos^4\left(kd\right)\phi_{\text{zpf},k}^4,\\
		\Lambda_k&=+\frac{E_{\text{S}}\left(\Phi_\text{DC}\right)}{720\hbar}\cos^6\left(kd\right)\phi_{\text{zpf},k}^6.
	\end{aligned}
	\right.
\end{equation}
The value of $kd$ should be determined from the constraint condition Eq. (\ref{constraint}).
Hereafter we will only consider the fundamental mode $k_0$ and hence will drop the subscripts of $\alpha_0$, $K_0$ and $\Lambda_0$ for convenience.

Still, for $\chi\ll1$, $\alpha$ can be approximated as
\begin{equation}
	\label{alpha}
	\begin{aligned}
		\alpha\approx-\frac{\pi\chi\omega_0 \Phi_\text{P}}{\Phi_{0}}\tan\left(\frac{\pi \Phi_\text{DC}}{\Phi_{0}} \right).
	\end{aligned}
\end{equation}

The values of the non-linearity coefficients, $K$ and $\Lambda$, are calculated as follows.

First, the Josephson coupling energy $E_\text{J}$ and the SQUID capacitance $C_\text{S}$, determined from the critical current $I_\text{c}\approx8\,\si{\micro\ampere}$ of a single JJ and its plasma frequency $\omega_\text{pl}/(2\pi)\approx80\,\si{\giga\hertz}$, are $E_\text{J}\approx2.633\times10^{-21}\,\si{\joule}$ and $C_\text{S}\approx192.4\,\si{\femto\farad}$, respectively. At the operational point for detection, the DC flux applied is  $\Phi_{\text{DC}}/\Phi_{0}\approx0.3618$.

Next, we extracted $L_\text{cav}\approx2.023\,\si{\nano\henry}$ (or equivalently, the bare resonance frequency $\omega_{\lambda/4}/(2\pi)\approx6.179\,\si{\giga\hertz}$) 
and $L_\text{S}(\Phi_{\text{DC}})\approx45.94\,\si{\pico\henry}$ from the fitting of the resonant frequency under DC-flux bias sweep, using the formula derived from Eq. (\ref{mode}) (and see also  Ref.~\cite{Krantz_2013})

\begin{equation}
	\begin{aligned}
		\omega_0(\Phi_{\text{DC}})
		= \frac{\omega_{\lambda/4}}{1+\chi(\Phi_{\text{DC}})}
		= \frac{\omega_{\lambda/4}}{1+L_\text{S}(\Phi_{\text{DC}})\left/L_\text{cav}\right.}.
	\end{aligned} \label{eq:omegazero}
\end{equation}

\begin{figure}[htb]
	\centering
	\includegraphics[width=.7\linewidth]{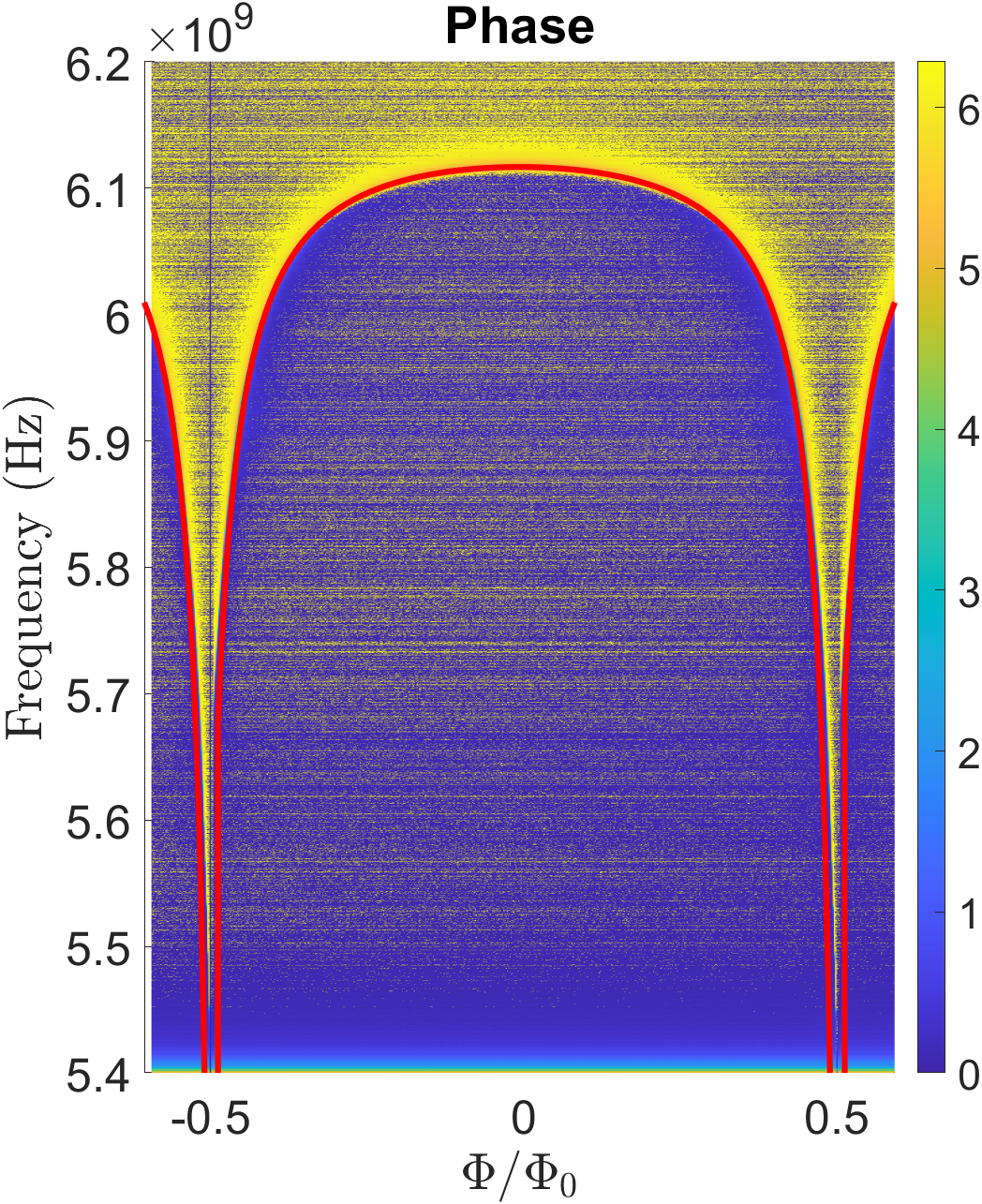}
	\caption{Fitting (the red line) for the DC sweep data of the K-JPA, with $\omega_{\lambda/4}/(2\pi)=6.179\,\si{\giga\hertz}$ -- see Eq. \ref{eq:omegazero}.}
	\label{DCsweep}
\end{figure}

To better fit the data, we have adopted the method used in \cite{pogorzalekFluxdrivenJosephsonParametric2017}, which takes the finite loop inductance ($L_\text{loop}=4.846\,\si{\pico\henry}$, obtained through Ansys HFSS simulation) of the SQUID into account, such that $L_\text{S}(\Phi_{\text{DC}})$ is corrected to
\begin{equation}
	L_\text{S}(\Phi_{\text{DC}})=\frac{\Phi_{0}}{4\pi I_\text{c}\cos\left[\phi_-^\text{min}\left(\Phi_{\text{DC}}\right)\right]},
\end{equation}
where $\phi_-^\text{min}\left(\Phi_{\text{DC}}\right)$ is determined by minimizing the dc SQUID potential \cite{pogorzalekFluxdrivenJosephsonParametric2017, clarkeSQUIDHandbookFundamentals2006}
\begin{equation}
	\frac{U(\phi_\text{S},\phi_-)}{2E_\text{J}}=
	\frac{1}{\pi\beta_\text{L}}\left(\phi_--\pi\frac{\Phi_\text{DC}}{\Phi_0}\right)^2
	-\cos\phi_\text{S}\cos\phi_-
	-\frac{I}{2I_\text{c}}\phi_\text{S},
\end{equation}
with negligible $\propto I$ term, $\phi_-\equiv\left(\phi_{\text{J},1}-\phi_{\text{J},2}\right)/2$ and the screening parameter $\beta_\text{L}\equiv2L_\text{loop}I_\text{c}/\Phi_0\approx0.0375$.
Figure \ref{DCsweep} shows the fitting result. From the fitting curve we extract the resonance frequency at zero flux bias $6.117\,\si{\giga\hertz}$ and the inductive participation ratio at the operational point $\chi_0\approx0.0227$.

Alternatively, the cavity capacitance and inductance can be determined based on the design, as $L_\text{cav}=L_ld\approx1.917\,\si{\nano\henry}$ and $C_\text{cav}=C_ld\approx744.6\,\si{\femto\farad}$.

Therefore, the fundamental mode is given by the approximated solution, Eq. (\ref{mode}), as $k_0d\approx1.536$, which means $\omega_0/(2\pi)\approx6.042\,\si{\giga\hertz}$, $C_0\approx413.9\,\si{\femto\farad}$ and $\phi_{\text{zpf},0}\approx0.1760$. For comparison, using the design parameters, the corresponding values are $k_0d\approx1.532$, $\omega_0/(2\pi)\approx6.454\,\si{\giga\hertz}$, $C_0\approx382.0\,\si{\femto\farad}$ and $\phi_{\text{zpf},0}\approx0.1773$.

Finally, plugging the values above into Eq. (\ref{nonlinear_coeff}), we obtain $K/(2\pi)\approx-0.2084\,\si{\kilo\hertz}$ and $\Lambda/(2\pi)\approx2.604\times10^{-4}\,\si{\hertz}$, comparing with $K/(2\pi)\approx-0.3115\,\si{\kilo\hertz}$ and $\Lambda/(2\pi)\approx4.911\times10^{-4}\,\si{\hertz}$ when the design values are used.
For convenience, we summarize these results in Table \ref{tab:circuit_parameters}.
\begin{table}
	\begin{tabular}[t]{c|cccccccc}
		\toprule[1pt]
		\multirow{2}{4em}{\centering Para- \\ meters}  & $L_\text{cav}$ & $C_\text{cav}$ & $k_0d$ & $\omega_0/2\pi$ & $C_0$ & $\phi_{\text{zpf},0}$ & $K/2\pi$ & $\Lambda/2\pi$ \\
		 & $(\si{\nano\henry})$  & $(\si{\femto\farad})$  &  & $(\si{\giga\hertz})$    &  $(\si{\femto\farad})$  &  & $(\si{\kilo\hertz})$  & $(\si{\milli\hertz})$ \\
		\midrule
		Fitting & 2.023 & 809.2 & 1.536 & 6.042 & 413.8 & 0.1760 & -0.208 & 0.260 \\
		Design & 1.917 & 744.6 & 1.532 & 6.454 & 381.8 & 0.1773 & -0.312 & 0.491 \\
		\bottomrule
	\end{tabular}
	\caption{Circuit parameters.}
	\label{tab:circuit_parameters}
\end{table}

\subsection*{Measurement setup}

\begin{figure}
	\adjincludegraphics[ width=0.97\linewidth]{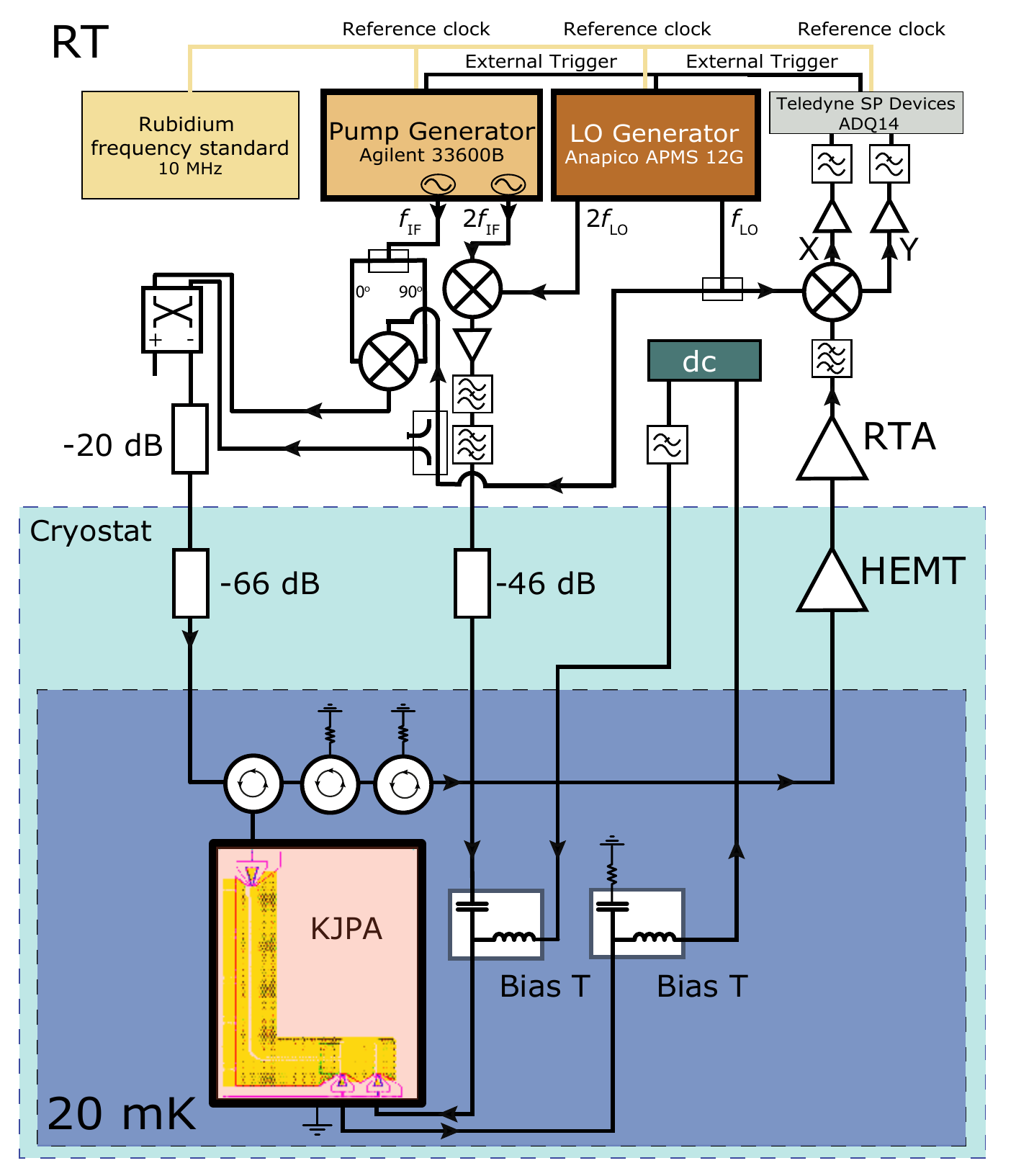}
	\caption{Illustration of experimental setup. The sample is cooled down to 20 mK in a BlueFors LD400 dry dilution refrigerator. See text for detailed information.}
	\label{fig:fullcircuitschematic}
\end{figure}

The full experimental setup, including the instruments at room temperature, is presented in Fig. 	\ref{fig:fullcircuitschematic}. The K-JPA chip is thermally anchored to the mixing chamber ($\approx20\,\si{\milli\kelvin}$) of a dry dilution cryostat. The chip has three ports: one is used for the signal  and the two ports are used together for DC biasing and flux pumping. A $10$-$\si{\mega\hertz}$ Rubidium frequency standard serves as the common reference clock. For the probe signal, the intermediate frequency ($f_\text{LO}=14\,\unit{\MHz}$) is combined with the local oscillator ($f_\text{LO}=6.028\,\unit{\GHz}$) using an IQ mixer, while for the pump the frequencies are doubled. To obtain a clean signal we use the technique of image band suppression with a 90-degree hybrid coupler. The same is achieved on the pump line by strong bandpass filtering using a combination of homemade bandpass cavities for $2f_\text{LO}$ and image band rejection after upconversion. The probe and pump are attenuated at different temperature stages ($-66$ and $-46$ dB respectively), and several circulators are used to provide isolation between input and reflected signals and to reduce back action. The reflected signal is amplified by a cryogenic high-electron-mobility transistor (HEMT) amplifier and a room-temperature amplifier (RTA), successively. The detection is done by a heterodyne arrangement, where the signal is downconverted to $f_\text{IF}=14\,\unit{\MHz}$ and digitized by the data acquisition card; after this it is multiplied by $\exp(-i2\pi f_\text{IF}t)$ in order to downcovert it to zero frequency. 

\begin{figure}
	\adjincludegraphics[ width=0.97\linewidth]{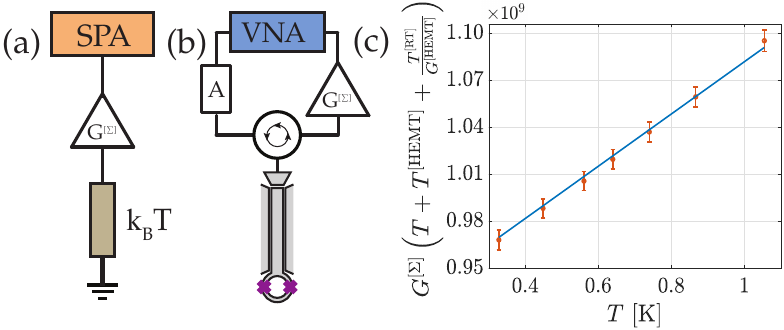}
	\caption{Calibration protocol. 
		(a) Simplified schematic of the circuits used for measuring the gain in the measurement chain. A 50 $\Omega$ load is gradually heated and its thermal noise power is amplified and measured at room temperature with a spectrum analyzer (SPA). 
		(b) Simplified schematic of the experimental setup for obtaining the attenuation coefficient $A$ from the "-" port hybrid coupler  (see  Fig. \ref{fig:fullcircuitschematic}). This is done by measuring the $S_\text{21}$ scattering parameter with a vector network analyzer (VNA).
		(c) Linear fitting of the measured power spectral density divided by $k_{\rm B}$ of the thermal noise at different temperatures of the 50 Ohm termination. 
	}
	\label{cal_scheme}
\end{figure}

\subsection*{Calibration of mean photon number in probe signals}
\label{sec:calibration-of-mean-photon-number-in-probe-signals}

For the calibration of the mean photon number we use a resistor realized as a 50 $\Omega$ termination in one of the circulators in Fig. \ref{fig:fullcircuitschematic}.
The calibration is done in two steps: measurement of the system gain with a spectrum analyzer, and measurement of attenuation in the input line with a vector network analyzer.

The system gain calibration procedure shown in Fig. \ref{cal_scheme}(a) consists of measuring the power spectrum density of the Johnson-Nyquist noise for different temperatures of a 50 $\Omega$ resistor. We use the formula for thermal noise power with equal source-load impedances $\mathbb{P}_\text{T}=\Delta f k_{\rm B} T $, from which the received power per unit of bandwidth can be written in terms of noise temperatures of source $T$ and of the amplifiers $T^{\text{[RT]}},T^{[\text{HEMT}]}$, multiplied by the system gain $G^{[\Sigma]}=G^{\text{[HEMT]}}G^{\text{[RT]}}$,

\begin{equation}
	\frac{\mathbb{P}_\text{T}}{\Delta f } = k_{\rm B} G^{[\Sigma]} \left( T+T^{\text{[HEMT]}}+\frac{T^{\text{[RT]}}}{G^{\text{[HEMT]}}}\right).
	\label{gain_cal}
\end{equation}

The term $T^{\text{[preamp]}}=T^{\text{[HEMT]}}+\frac{T^{\text{[RT]}}}{G^{\text{[HEMT]}}}=5.2\pm0.1$K is the combined noise temperature of amplifiers, where the largest contribution comes from the HEMT amplifier.
The total system gain is extracted from the linear interpolation with temperature, as shown in Fig. \ref{cal_scheme}(c). It also includes frequency transition losses/amplification in the spectrum analyzer circuit part. We obtain $G^{[\Sigma ]}= 82.4\pm0.2$ dB, where the error is associated to a single calibration.

To obtain the attenuation $A$, we replace in the input line the noise generated by the resistor with the actual probe field and we use a vector network analyzed to find the transmission coefficient $S_\text{21}$, see  Fig. \ref{cal_scheme}(b). This is done in continuous-wave operation at a frequency 6.042 GHz, where the cavity is far-detuned. Then, the total attenuation provided by all elements in the probe line (from the output represented as "-" of the hybrid coupler on Fig. \ref{fig:fullcircuitschematic} to cavity) can be calculated as $A=S_\text{21}-G^{[\Sigma ]}$. 

Such calibration scheme assumes lossless connections to the sample and it neglects the losses in the circulators; however, these can be estimated and included as a  small systematic error correction. We obtain $A=-97.2 \pm 0.2$\,dB. 
This figure includes also the room temperature $-20$\,dB attenuator shown in Fig.~\ref{fig:fullcircuitschematic}.
Room-temperature and cryogenic cables as well as connectors amount for an additional 11 dB of attenuation. Finally, the probe power $\mathbb{P}$ reaching the sample is obtained by multiplying the attenuation $A$ with the input power at room temperature. The latest is calibrated with the SPA, using
the full setup including mixers, LO cancellation and filters.

The average photon number in a pulse $\bar{n}$ is calculated from the probe power $\mathbb{P}$
(corresponding to an average energy per pulse $\hbar \omega \bar{n} = \mathbb{P}\tau$) as 
\begin{equation}
	\bar{n}=\frac{\tau\mathbb{P}}{\hbar \omega}.
\end{equation}

To obtain the point-to-point error in $\bar{n}$ we consider three sources of error: errors in $\mathbb{P}$, errors in $A$, and errors 
caused by reconnection cables from calibration to measurement. 
We estimate each of these errors to a maximum of $\sigma^{\rm (log)} (\text{dB}) \approx 0.2$ dB, where we use the convention that the dB error in a power-like quantity $A$ is $\sigma_{A}^{\rm (log)}(\mathrm{dB}) = 10 \log (1 \pm \sigma_A /A)$. For small errors the $\pm$ sign does not make a significant difference.
Then the accumulated error can be calculated from the standard formula of error propagation
\begin{equation}
	\frac{\sigma_{ABC\dots}}{ABC\dots}=\sqrt{\frac{\sigma_{A}^2}{A^2}+ \frac{\sigma_{B}^2}{B^2} + \frac{\sigma_{C}^2}{C^2} + \dots},
\end{equation}
so in our case $\sigma_{\bar{n}}=\sqrt{3}\sigma$
and we obtain $\sigma^{\rm (log)}_{\bar{n}}(\text{dB}) \approx 0.35$ dB.

\subsection*{Operation as detector}

To operate the device as a detector, a sequence of pulses and operations is applied as presented in Fig.~\ref{waveforms}. First, the system is brought from the starting point (SP) to the operational point (OP) by applying the pump of amplitude $\alpha$ for a time $\tau_{\rm P}$. Then the probe pulse $b=|b|\exp (-i \varphi )$ is applied, with a fixed delay of $0.228\,\unit{\micro\second}$ with respect to the pump. This delay is chosen such that the probe and the pulse arrive at the sample with the phase difference that maximizes the switching and also to ensure that the ringing is quenched.

In our experiments the frequency of the probe field is chosen equal to the half of pump frequency $\omega/(2\pi) = \omega_\text{P}/(4\pi) = 6.042\,\unit{\GHz}$. 
The power can be expressed as $\mathbb{P} = \hbar \omega|b|^2$, with $\bar{n} = |b|^2 \tau$: a power $\mathbb{P} = -174~{\rm dBW} = -144~{\rm dBm}
\approx 4.0 \times 10^{-18}$ W corresponds to $|b|\approx 1 \times 10^3 ~\sqrt{\rm Hz}$. Finally, the readout operation is performed during a time-window $\tau_{R}$.

Finally, all the parameters presented above are subjected to experimental errors. We have done a thorough estimation of these errors and their origin, see Table \ref{tab:params}.

\begin{widetext}
\begin{table*}[t]
	\centering
	\begin{tabular}{|l|c|c|l|}
		\hline
		\parbox[t]{3.5cm}{\textbf{Parameter}} & \textbf{Value} & \textbf{Error} &\parbox[t]{10.5cm}{ \textbf{Description of error}} \\
		\hline
		\parbox[t]{3.5cm}{$\omega /(2\pi)$} & $6.042~\si{\giga\hertz}$ & - & \parbox[t]{10.5cm}{-} \\
		\hline
		\parbox[t]{3.5cm}{$\kappa/(2\pi)$} & $4.44~\si{\mega\hertz}$ & $\pm10~\si{\kilo\hertz}$ & \parbox[t]{10.5cm}{Extracted from S21 fitting of unpumped resonator, defined by both the precision of the measurement and the fitting. }\\
		\hline
		\parbox[t]{3.5cm}{$\gamma/(2\pi)$} & $2.30~\si{\mega\hertz}$ & $\pm10~\si{\kilo\hertz}$ & \parbox[t]{10.5cm}{Extracted from S21 fitting of unpumped resonator, defined by both the precision of the measurement and the fitting.} \\
		\hline
		\parbox[t]{3.5cm}{Resonance frequency or \newline detuning at OP, $\Delta/(2\pi$)} & $0.7~\si{\mega\hertz}$ &  $\pm 50~\si{k\hertz}$ & \parbox[t]{10.5cm}{Originates mostly from fitting of DC sweep data, fitting of the phase diagram, and defining the critical threshold at zero detuning. }\\
		\hline
		\parbox[t]{3.5cm}{Kerr non-linearity \newline $K/(2\pi$) (experimental)} & $-0.2~\si{\kilo\hertz}$ & $\pm 20\%$ &\parbox[t]{10.5cm}{ Extracted by fitting with the Langevin equation and measuring the gain coefficient, squeezing and entanglement. }\\
		\hline
		\parbox[t]{3.5cm}{Kerr non-linearity \newline $K/(2\pi$) (calculation)} & $-0.21~\si{\kilo\hertz}$ & $\pm 0.1~\si{\kilo\hertz}$ & \parbox[t]{10.5cm}{Error comes from critical current ($I_{\mathrm{c}}$) of a JJ and fitting of the DC sweep data used to extract $L_{\mathrm{cav}}$ and $L_{\mathrm{S}}(\Phi_{\mathrm{DC}})$.} \\
		\hline
		\parbox[t]{3.5cm}{Pump strength $\alpha/(\kappa+\gamma)$ for $1$-$\si{\micro\second}$ experiment} & $0.51$ & $\pm 0.004$ & \parbox[t]{10.5cm}{Errors come from finding the location of the critical threshold in the phase diagram and its resolution.} \\
		\hline
		\parbox[t]{3.5cm}{Back-forth \newline propagation delay} & $250~\si{\nano\second}$ & $\pm20~\si{\nano\second}$ & \parbox[t]{10.5cm}{Sampling rate. $\pm2$ samples included.} \\
		\hline
		\parbox[t]{3.5cm}{Pulse duration \newline as seen by the sample} & varies & $\pm100~\si{\nano\second}$ & \parbox[t]{10.5cm}{$\sim 100$ ns cavity ringing time.} \\
		\hline
		\parbox[t]{3.5cm}{Pump-probe \newline delay (from generator)} & $228~\si{\nano\second}$ & $\pm 2~\si{\nano\second}$ &\parbox[t]{10.5cm}{Distance between 2 generator's samples is $4~\si{\nano\second}$.} \\
		\hline
		\parbox[t]{3.5cm}{Probe power coupled into the device} & varies & $\pm 0.85~\si{\decibel}$ & \parbox[t]{10.5cm}{Calibration errors add up to $0.35~\si{\decibel}$ (includes power, gain, attenuation, and reconnection errors). Systematic errors from drifts, connectors, sample design and package, etc.  may add up to a maximum of $0.5~\si{\decibel}$. For example: an overall factor $1.1$ rescaling in b is equivalent to a $0.83~\si{\decibel}$ addition in power.}
		\\
		\hline
	\end{tabular}
	\caption{Description of parameters and associated errors.}
	\label{tab:params}
\end{table*}
\end{widetext}

\section*{Appendix B: Theoretical model for phase transitions}

To understand the phase diagram presented in Fig. \ref{QPTMeanVar}, we examine the dynamics generated by the general Hamiltonian for the parametric oscillator with a Kerr nonlinearity $K$ and with the pump strength $\alpha = |\alpha|\exp(-i \theta_{\rm P})$,
\begin{equation}
		H_{\rm sys} = \hbar\omega_{\rm 0} a^{\dag}a + \frac{\hbar}{2}\left[ \alpha e^{-i \omega_{\rm P}t} + c.c.\right](a+a^{\dag})^2 + \hbar K (a + a^\dag)^4 .
\end{equation}
By using the following definition for the quadratures
\begin{equation}
\mathcal{Q}=\frac{1}{\sqrt{2}}(a + a^{\dag})~{\rm and}~
\mathcal{P}=\frac{1}{i\sqrt{2}}(a - a^{\dag}),
\end{equation}   
satisfying the Heisenberg commutation relations $[\mathcal{Q}, \mathcal{P}] = i$, we obtain
\begin{equation}
	H_{\rm sys} = \frac{\hbar \omega_{0}}{2} \left(\mathcal{P}^2 + \mathcal{Q}^2\right) + 2 \hbar |\alpha| \mathcal{Q}^2 \cos(\omega_{\rm P}t + \theta_{\rm P}) + 4\hbar K \mathcal{Q}^4
\end{equation}
(up to a constant value $\hbar \omega_0/2$).
We can transform the Hamiltonian in a rotating frame by the unitary operator $U = \exp [-i(\omega_{\rm P}t/2 + \theta_{\rm P} /2 + \pi/4)a^{\dag}a]$ \cite{lin2015critical}, obtaining a new Hamiltonian $U^{\dagger}H_{\rm sys}U -i \hbar U^{\dagger}\dot{U}$. Under the rotating wave approximation (RWA) this Hamiltonian can be further simplified to 
\begin{equation}
	H_{\rm sys}^{\rm (RWA)} = \frac{\hbar \Delta}{2}(\mathcal{P}^2 + \mathcal{Q}^2)
	+ \frac{\hbar |\alpha |}{2} (\mathcal{P}\mathcal{Q} +
	\mathcal{Q}\mathcal{P} ) + \frac{3\hbar K}{2} (\mathcal{P}^2 + \mathcal{Q}^2)^2,
\end{equation}
where $\Delta \equiv \omega_{0}-\omega_{\text{P}}/2$, yielding the corresponding 
Heisenberg-Langevin equations for the $\mathcal{Q}$ and $\mathcal{P}$ variables
\begin{eqnarray}
\dot{\mathcal{Q}} &=& \left[|\alpha | - \alpha_{\rm c}(0)\right] \mathcal{Q} + \Delta \mathcal{P} +  \nonumber \\
 & &+ 6 K \left[ \frac{1}{2}(\mathcal{P}\mathcal{Q}^2 + \mathcal{Q}^2\mathcal{P}) + \mathcal{P}^3\right] - \sqrt{\kappa + \gamma} \xi_{\mathcal{Q}_{\rm in}} , \label{eq:dotq}\\
\dot{\mathcal{P}} &=& -\left[|\alpha | + \alpha_{\rm c}(0) \right] \mathcal{P} - \Delta  \mathcal{Q} - \nonumber \\
& & - 6 K \left[ \frac{1}{2}(\mathcal{Q}\mathcal{P}^2 + \mathcal{P}^2\mathcal{Q}) + \mathcal{Q}^3\right] -  \sqrt{\kappa + \gamma} \xi_{\mathcal{P}_{\rm in}} . \label{eq:dotp}
\end{eqnarray}
Here the last terms $\xi_{\mathcal{Q}_{\rm in}}$ and $\xi_{\mathcal{P}_{\rm in}}$ are total quadrature input noises, while $\alpha_{\rm c} (0 ) = (\kappa + \gamma )/2$ is the pump amplitude corresponding to the critical point at $\Delta =0$ observed in the phase diagram. Indeed, considering the linearization of Eqs. (\ref{eq:dotq}, \ref{eq:dotp}) and with the notation $\alpha_{\rm c} (\Delta )= \sqrt{\Delta^2 + (\kappa + \gamma)^2/4}$ we notice that 
 \begin{equation}
 	\alpha_{\rm c} (\Delta )^2 - |\alpha|^2
 \end{equation}
is the determinant whose zeroes yield the separation between phases in Fig. \ref{QPTMeanVar}.
For values of $|\alpha |$ and $\Delta$ not too far from the transition, we see from the equations above that the time scales of the dynamics separate: $\mathcal{Q}$ becomes a slow variable and $\mathcal{P}$ a fast variable \cite{lin2015critical, marthaler2007quantum, peano2012sharp,dykmanFluctuatingNonlinearOscillators2012}.  As a result, $\mathcal{P}$ will follow 
$\mathcal{Q}$ adiabatically, and by imposing 
$\dot{\mathcal{P}}=0$
we can obtain this adiabatic value of $\mathcal{P}$ from Eq. (\ref{eq:dotp})
and use it in Eq. (\ref{eq:dotq}). After neglecting some relatively small terms, we obtain the equation of motion of $\mathcal{Q}$ in terms of an effective potential 
$\mathcal{U}(\mathcal{Q})$
\begin{equation}
\dot{\mathcal{Q}} =	- \partial_{\mathcal{Q}}\mathcal{U}(\mathcal{Q}),
\end{equation}
with 
\begin{eqnarray}
		\mathcal{U}(\mathcal{Q})  &=&  \frac{2}{|\alpha|  + \alpha_{\rm c}(0)} \times \label{eq:eff} \\
		& & \times  \left[ \left(\alpha_{\rm c}(\Delta)^2 - |\alpha|^2 \right)\frac{\mathcal{Q}^2}{4} + 3\Delta K \frac{\mathcal{Q}^4}{2} + 3 K^2 \mathcal{Q}^6 \right] .\nonumber 
\end{eqnarray}	
This effective potential explains fully the phases observed experimentally, see Fig. \ref{QPTMeanVar}.

One can immediately check that $\mathcal{Q}_{0}=0$ is always a local extremum, irrespective to the values of $|\alpha|$ and $\Delta$. It is a maximum if $|\alpha | > \alpha_{\rm c}(\Delta)$ and a minimum otherwise. If $|\alpha| > \alpha_{\rm c}(0)$ for $\Delta > 0$, or if 
$|\alpha | > \alpha_{\rm c}(\Delta)$ for  $\Delta < 0$ we have two local minima at  $\pm \mathcal{Q}_{\rm min}$
\begin{equation}
	\mathcal{Q}_{\rm min} = \frac{1}{\sqrt{6|K|}}\sqrt{\Delta + \sqrt{|\alpha|^2 - \alpha_{\rm c}(0)^2}}.
\end{equation}
In addition, for $\Delta > 0$ and if $|\alpha| > \alpha_{\rm c}(0)$ and $|\alpha| < \alpha_{\rm c} (\Delta )$ we get two maxima at $\pm \mathcal{Q}_{\rm max}$, where 
 \begin{equation}
 	\mathcal{Q}_{\rm max} = \frac{1}{\sqrt{6|K|}}\sqrt{\Delta - \sqrt{|\alpha|^2 - \alpha_{\rm c}(0)^2}}.
 \end{equation}
Thus, depending on the detuning and pump strength, the system can be localized in either the central well (so-called zero-amplitude state) or in one of the wells at $\mathcal{Q}_{\rm min}$ (so-called oscillatory states), see Fig. \ref{QPTMeanVar}.

\section*{Appendix C: Quantum efficiency}

Here we give a brief overview of the general quantum theory of non-number-resolving detectors. These detectors -- also called bucket detectors -- are common in quantum optics, and they have two states, denoted here by $\texttt{0}$ and $\texttt{1\raisebox{-0.5ex}{+}}$. In this framework, the detection can be regarded as a form of hypothesis testing \cite{hypothesis2021, PhysRevE.85.016708}.
 The associated POVM operators are \cite{Kok2000,RevModPhys2007,migdallSinglephotonGenerationDetection2013}
\begin{eqnarray}
	\Pi_{\texttt{0}} &=& \sum_{n=0}^{\infty}(1-\eta )^{n}|n\rangle\langle n| ,\\
	\Pi_{\texttt{1\raisebox{-0.5ex}{+}}} &=& \mathbb{I} - \Pi_{0} ,
\end{eqnarray}
resulting in probabilities $p_{\texttt{0}}= {\rm Tr}\{\Pi_{\texttt{0}} \rho\} = \sum_{n=0}^{\infty}(1-\eta )^{n} P_{n}$ and $p_{\texttt{1\raisebox{-0.5ex}{+}}}= {\rm Tr}\{\Pi_{\texttt{1\raisebox{-0.5ex}{+}}} \rho\}= 1 - p_{\texttt{0}}$, where $\rho$ is the state at the input of the detector with photon number probabilities $P_{n} = {\rm Tr}\{ |n\rangle\langle n| \rho \} $
\cite{Kok2000,RevModPhys2007,migdallSinglephotonGenerationDetection2013}. The efficiency $\eta$ is  identical to the probability  $p_{\texttt{1\raisebox{-0.5ex}{+}}}$ of detection when $\rho = |1\rangle \langle 1|$ is a Fock state consisting of a single photon.

If a coherent state 
$\rho_{\bar{n}} = \exp (-\bar{n})\sum_{m,n=0}^{\infty}
\frac{\bar{n}^{(n+m)/2}}{\sqrt{n!}\sqrt{m!}}|n\rangle \langle m|$ 
with $\bar{n}$ average number of photons is present at the input, we have 
$p_{\texttt{0}} = \exp (-\eta \bar{n})$ and $p_{\texttt{1\raisebox{-0.5ex}{+}}} = 1-p_{0}$. This reproduces exactly the Poissonian statistics
with number-of-photon probabilities $P_{n} =\frac{\bar{n}^n}{n!}e^{-\bar{n}}$ of the coherent field. Also the single-photon detection efficiency can be related to the coherent-state detection probability at 
$\bar{n}=1$;  we have $\eta = -\ln \left(1-p_{\texttt{1\raisebox{-0.5ex}{+}}}\vert_{\bar{n}=1}\right)$. The maximum 
$p_{\texttt{1\raisebox{-0.5ex}{+}}}$ is obtained for
$\eta =1$, namely ${\rm max}\{p_{\texttt{1\raisebox{-0.5ex}{+}}}\vert_{\bar{n}=1} \} = 1 - e^{-1} = 0.632$.

The next step is to introduce the dark count events, which occur with probability  $p_\texttt{dark}$. We can obtain the probability that the detector does not click $p_{\texttt{0$\land\neg$dark}}$ as the product of the probability $p_{\texttt{0}}$ 
and the probability $p_{\neg\texttt{dark}} = 1-p_\texttt{dark}$ of no dark counts,
$p_{\texttt{0$\land\neg$dark}}=(1-p_\texttt{dark})p_{\texttt{0}}$, assuming that the two processes are independent of each other \cite{leePhotostatisticsPhotonnumberDiscriminating2004}.

Then, the probability of a detection event (resulting from either an input photon or a dark click) is 
$p_{\texttt{1\raisebox{-0.5ex}{+}$\lor$dark}}=1-p_\texttt{0$\land\neg$dark} = p_{\texttt{1\raisebox{-0.5ex}{+}}}+p_\texttt{dark}-p_{\texttt{1\raisebox{-0.5ex}{+}}} p_\texttt{dark}$, therefore
\begin{equation}
	p_{\texttt{1\raisebox{-0.5ex}{+}}}=\frac{p_{\texttt{1\raisebox{-0.5ex}{+}$\lor$dark}}-p_\texttt{dark}}{1-p_\texttt{dark}}.
	\label{p1}
\end{equation} 
On the other hand, for a coherent state with $\bar{n}$ average number of photons,
we have $p_{\texttt{0}} = \exp (-\eta \bar{n})$ and $p_{\texttt{1\raisebox{-0.5ex}{+}}} = 1-p_{0}$, or
\begin{equation}
	p_{\texttt{1\raisebox{-0.5ex}{+}}}=1-p_\texttt{0}=1-\exp[-\bar{n}\eta],
	\label{p1_exp}
\end{equation}
and as a result we find

\begin{equation}
	\eta =\frac{1}{\bar{n}}\ln\frac{1-p_\texttt{dark}}{1-p_{\texttt{1\raisebox{-0.5ex}{+}$\lor$dark}}}.
\end{equation}
Another way to obtain this relation is by combining
$p_\texttt{0$\land\neg$dark} = p_{\texttt{0}}p_\texttt{$\neg$dark}$,
and $p_{\texttt{0}} = \exp (-\eta \bar{n})$.

{\it Noise equivalent power.}
A convenient measure of sensitivity typically used in single-photon detection and bolometry is the noise-equivalent power \cite{Komiyama5475208,Cleland1992,astafievSinglePhotonDetectionQuantum2003}. 
Consider a switching process as above, where the events happen randomly such that the probability that there is a switching event in a short time interval $dt$ is $\Gamma dt$. We define a generic random signal associated with switching events at times $t_j$ as $r(t) = \sum_{j} \delta (t-t_j)$ and its fluctuations around the ensemble average $r_{n}(t) = r(t) - \langle r(t) \rangle $. The double-sided spectral density of $r_n$ at a frequency $f$ is in this case $S_{r_{n}r_{n}} (f)= \Gamma$, and  the single-sided spectral density is twice this value. From the Wiener-Khinchin theorem we obtain the temporal mean square $\overline{r_{n}^2} = \int_{0}^{\infty} df S_{r_{n}r_{n}} (f)$ and since the detection is in a finite bandwidth B we obtain $\overline{r_{n}^2} = B \Gamma$. The noise-equivalent power (NEP) is defined by considering an equivalent input signal that would produce the noise $r_{n}$. The power of this signal should be  $\mathbb{P}_{n} = \frac{1}{\eta} r_{n} hf$ -- higher than just $r_{n}hf$ by a factor of $1/\eta$ to compensate for the limited 
efficiency of the detector. Then, NEP is defined as ${\rm NEP} = \sqrt{\overline{\mathbb{P}_{n}^2}/B}$ and therefore  
\begin{equation}
	\mathrm{NEP} = \frac{1}{\eta}\sqrt{2\Gamma} hf.
\end{equation}

\section*{Appendix D: Switching rates}

The noise terms in Eqs. (\ref{eq:dotq}, \ref{eq:dotp}) originate from the thermal fluctuations of the electromagnetic field in the input line, with correlations
\begin{eqnarray}
	\langle  \xi_{\mathcal{Q}{\rm in}}(t)  \xi_{\mathcal{Q}_{\rm in}}(t')  \rangle &=& (\bar{n}_{T}+ 1/2) \delta (t-t'), \nonumber \\
	\langle  \xi_{\mathcal{P}_{\rm in}}(t)  \xi_{\mathcal{P}_{\rm in}}(t')  \rangle &=& 
	(\bar{n}_{T}+ 1/2) \delta (t-t') , \nonumber \\
	\langle  \left[\xi_{\mathcal{Q}_{\rm in}}(t),  \xi_{\mathcal{P}_{\rm ib}}(t') \right] \rangle &=& i \delta (t - t') . \nonumber
\end{eqnarray}
As a result, the system prepared in the zero-amplitude state at $Q=0$ may switch either to the left or to the right wells located around  $\pm \mathcal{Q}_{\rm min}$. 
This manifests in the measurements as dark counts.

The presence of a coherent field $b=|b|\exp (-i \varphi )$ at the input can be modeled by adding the probe Hamiltonian
\begin{equation}
	-2\sqrt{2}\hbar \sqrt{\kappa }|b| \cos \left(\frac{\omega_\text{P} t}{2} + \varphi \right) \mathcal{P}.	
\end{equation}
The field contains on average $\bar{n} = |b|^2 \tau$ in the time interval $\tau$.
We use again the same rotating frame $U$ defined by half the pump frequency  and we apply the rotating wave approximation, obtaining a new set of Heisenberg-Langevin equations
\begin{align}
	\dot{\mathcal{Q}} =& \left[|\alpha | - \alpha_{\rm c}(0) \right] \mathcal{Q} + \Delta \mathcal{P} + 6 K \left[ \frac{\mathcal{P}\mathcal{Q}^2 + \mathcal{Q}^2\mathcal{P}}{2} + \mathcal{P}^3\right]  \nonumber \\
	& -\sqrt{2\kappa} |b|\cos\left(\frac{\theta_{\rm P}}{2} +\frac{\pi}{4}
	-\varphi \right)  - \sqrt{\kappa + \gamma} \xi_{\mathcal{Q}_{\rm in}} ,
	\label{eq:probedotq}\\
	\dot{\mathcal{P}} =& -\left[|\alpha | + \alpha_{\rm c}(0)\right] 
	\mathcal{P} - \Delta \mathcal{Q}  - 6 K \left[ \frac{\mathcal{Q}\mathcal{P}^2 + \mathcal{P}^2\mathcal{Q}}{2} + \mathcal{Q}^3\right] \nonumber \\ 
	& -\sqrt{2\kappa} |b|\sin\left(\frac{\theta_{\rm P}}{2} +\frac{\pi}{4}-\varphi \right)
	- \sqrt{\kappa + \gamma} \xi_{\mathcal{P}_{\rm in}} . \label{eq:probedotp}
\end{align}
The equations are invariant under $\varphi \rightarrow \varphi + 2 \pi$. The maximum switching rate is obtained by maximizing the effect of $|b|$ in the slow (and larger) variable $\mathcal{Q}$, therefore the phase that achieves this is $\varphi = \theta_{\rm P} /2 +\pi /4$. By following the same procedure as before, we can obtain a modified effective potential for the slow variable as
\begin{equation}
\mathcal{U}_{b} (\mathcal {Q}) =  \mathcal{U} (\mathcal {Q}) + \sqrt{2 \kappa}|b|\mathcal{Q} .\label{eq:Upotential}
\end{equation}
From here we see that the effect of the probe field is to tilt the potential $\mathcal{U} (\mathcal {Q})$, which is reminiscent of the way in which a current bias applied to a Josephson junction produces a washboard potential. The minima around $Q=0$ is only mildly affected by this. Most of the effect of the probe field on the potential amounts to a lowering of the left barrier to approximately $\mathcal{U}(\mathcal{Q}_{\rm max}) - \sqrt{2 \kappa} |b| \mathcal{Q}_{\rm max}$ and raising the right barrier to approximately $\mathcal{U}(\mathcal{Q}_{\rm max}) + \sqrt{2 \kappa} |b| \mathcal{Q}_{\rm max}$.

Next, we construct the associated Fokker-Planck equation for the probability density $W_{b}(\mathcal{Q},t)$,
\begin{equation}
	\partial_{t} W_{b} = \partial_{\mathcal{Q}} \left( W_{b} \partial_{\mathcal{Q}}  U_{b} \right) + D \partial_{\mathcal{Q}}^{2}W_{b},  \label{eq: FP}
\end{equation}	
where the diffusion is $D = (\kappa /2 + \gamma /2)(\bar{n}_{T}+ 1/2)$. We simulate numerically this equation with initial conditions corresponding to the starting point ($\alpha =0$) of each detection sequence. 

To obtain the dark count rate we observe that the probability that no switching event has occured in a time $t$ in the absence of an input field is given by $\int_{-\mathcal{Q}_{\rm th}}^{\mathcal{Q}_{\rm th}} W_{b=0}(\mathcal{Q},t)  d \mathcal{Q}$. This integral is calculated from the numerical solution of the Fokker-Planck equation Eq. (\ref{eq: FP}) for $b=0$. We then fit $\ln [\int_{-\mathcal{Q}_{\rm th}}^{\mathcal{Q}_{\rm th}} W_{b=0}(\mathcal{Q},t)  d \mathcal{Q}]$ with the linear function $-\Gamma_{\texttt{dark}} t+ \mathrm{const.}$, from which we get $\Gamma_{\texttt{dark}} = 167~\mathrm{kHz}$.
The dark count probability is obtained in the same way, by integrating 
Eq. (\ref{eq: FP}) over the sensing time (which coincides with the action of the pump)
\begin{equation}
	p_\texttt{$\neg$dark} = 1 -  p_{\texttt{dark}}= 
	\int_{-\mathcal{Q}_{\rm th}}^{\mathcal{Q}_{\rm th}} W_{b=0}(\mathcal{Q},\tau_{\rm P})  d \mathcal{Q} . \label{eq:darkcount}
\end{equation}
Similarly, when the probe field is on, we can find the probability that no switching event has occurred during $\tau_{\rm P}$ as 
 \begin{equation}
 	P_\texttt{0} =	p_\texttt{0$\land\neg$dark} = 1 -  p_{\texttt{1\raisebox{-0.5ex}{+}$\lor$dark}}= 
\int_{-\mathcal{Q}_{\rm th}}^{\mathcal{Q}_{\rm th}} W_{b}(\mathcal{Q},\tau_{\rm P})  d \mathcal{Q} ,
 \end{equation}
where $\mathcal{Q}_{\rm th}$ is the threshold value. The efficiency can be obtained immediately by using 
$p_\texttt{0$\land\neg$dark} = p_{\texttt{0}}p_\texttt{$\neg$dark}$,
 as explained also in the main text.

\bibliography{./SPDetector_final_refs.bib}

\end{document}